\documentclass[preprint,12pt]{elsarticle}
\usepackage[export]{adjustbox}
\usepackage{wrapfig}
\usepackage{lscape}
\usepackage[utf8]{inputenc} 
\usepackage{multirow}
\usepackage{booktabs}
\usepackage{subcaption}
\usepackage{diagbox}
\usepackage{slashbox}
\usepackage{float}
\usepackage{amsmath}
\usepackage{latexsym}
\usepackage{mathrsfs}
\usepackage{amsmath}
\usepackage{amssymb}
\usepackage{graphicx}
\usepackage{mathrsfs}
\usepackage[colorlinks=true, allcolors=blue]{hyperref}
\usepackage[numbers]{natbib} 
\usepackage[pdftex,dvipsnames]{xcolor}
\usepackage[colorinlistoftodos,prependcaption,textsize=tiny]{todonotes}

\newtheorem{definition}{Definition}
\newtheorem{example}{Example}
\newtheorem{proposition}{Proposition}

\journal{Engineering Applications of Artificial Intelligence}

\newcommand{\modif}[1]{\textcolor{black}{#1}}
\newcommand{\nzm}[1]{\textcolor{black}{#1}}

\begin{document}
\begin{frontmatter}
\title{An Online A/B Testing Decision Support System for Web Usability Assessment Based on a Linguistic Decision-making Methodology: Case of Study a Virtual Learning Environment}

\author[label0]{Noe Zermeño}
\author[label1]{Cristina Zuheros}
\author[label3]{Lucas Daniel Del Rosso Calache}
\author[label1]{Francisco Herrera}
\author[label2]{Rosana Montes\corref{cor1}}
\cortext[cor1]{Corresponding author.\\
E-mail address: rosana@ugr.es (R. Montes)}

\address[label0]{University of Guadalajara, Mexico.}
\address[label2]{Department of Software Engineering, Andalusian Research Institute in Data Science and Computational Intelligence (DaSCI), University of Granada, 18071, Granada, Spain.}
\address[label1]{Department of Computer Science and Artificial Intelligence, Andalusian Research Institute in Data Science and Computational Intelligence (DaSCI), University of Granada, 18071, Granada, Spain.}
\address[label3]{Sao Paulo State University (Unesp), School of Engineering, Campus of Sao Joao da Boa Vista, São Paulo, Brazil.}

\markboth{R. Montes}{An Online A/B Testing Decision Support System }

\begin{abstract}
In recent years, attention has increasingly focused on enhancing user satisfaction with user interfaces, spanning both mobile applications and websites. One fundamental aspect of human-machine interaction is the concept of web usability. In order to assess web usability, the A/B testing technique enables the comparison of data between two designs. Expanding the scope of tests to include the designs being evaluated, in conjunction with the involvement of both real and fictional users, presents a challenge for which few online tools offer support. We propose a methodology for web usability evaluation based on user-centered approaches such as design thinking and linguistic decision-making, named Linguistic Decision-Making for Web Usability Evaluation. This engages people in role-playing scenarios and conducts a number of usability tests, including the widely recognized System Usability Scale. We incorporate the methodology into a decision support system based on A/B testing. We use real users in a case study to assess three Moodle platforms at the University of Guadalajara, Mexico. 
\end{abstract}


\end{frontmatter}

\section{Introduction}

With the increasing use of websites and online platforms ---such as learning management systems (LMS) in the educational context--- there is a growing need to enhance the quality of software usability to ensure satisfactory user experiences (UX). Software engineers specializing in UX methodologies design and evaluate interfaces using user-centered techniques, such as \emph{Design for All}. Web accessibility has been in development for decades \cite{chisholm2001web} and in this regard, there are many guidelines for system engineers to follow and these include ISO standards (ISO9216-11~\cite{iso1998iec}, ISO25000~\cite{iso2011iso}) and also WCAG 2.2. guidelines from the World Wide Web Consortium (W3C)~\cite{wcag22}. In spite of this, however, there is a lack of clear guidance on how to validate web usability compliance. 

Web usability has been explored under various evaluation methods, such as expert-driven inquiry, inspection and testing. However, none of these methods has become a standard since they integrate a multitude of different factors in addition to the end-user’s own opinions. Furthermore, no software tool exists to offer support in every stage of the process (either by means of a single test or various tests, which are performed either by experts alone, or experts in conjunction with \textit{personas}~\cite{bookUX12}, or actual end-users). A standardized and cost-effective solution is to apply the system usability scale (SUS) questionnaire~\cite{Brooke:1996}. Despite the fact SUS questionnaire is widely used for usability assessment, on its own it does not capture the areas of opportunity in interface design required by different varieties of users~\cite{sauro2016quantifying,sauro2011practical}. This limitation is because SUS provides a general overview of usability, but does not address the specific needs of users with different roles or disabilities. \modif{Therefore, we suggest to complement it with user-centered design techniques.} 

The idea behind the Design Thinking (DT) methodology is to understand user needs, generate creative solutions, and rapidly prototype new ideas ~\cite{lockwood2010design}. \modif{It fosters empathy with users through techniques such as \textit{role-playing} (where testers assume the roles of different user profiles), encourage experimentation and ultimately deliver products, services, or solutions that address real-world problems. Designing the diversity of abilities of users from the earliest stages of the design process is essential for achieving improved usability.}

Another approximation for fostering a user-centered design is to provide inputs and outputs close to human reasoning. Computing with words (CW)~\cite{Martinez2010} is a methodology which operates with people's perceptions rather than numerical measures, resulting in flexibility in the interpretation of the results since they are expressed in natural language and not by numbers. \modif{This approach is useful in complex group decision-making scenarios, allowing experts to express their assessments in linguistic terms that incorporate uncertainty. In this line, Dong et al. \cite{dong2025integrated} developed a group decision-making method using probabilistic linguistic assessments for hotel site selection, demonstrating the effectiveness of capturing uncertainty in human preferences when making collective decisions.} 

If we consider usability assessment as the comparison of two or more versions of the same site, or even different related sites (alternatives) in relation to a set of attributes (or criteria), we can consider this to be a \modif{multi-criteria decision-making (McDM)} problem~\cite{libro2000}. \modif{McDM based on internal rough-fuzzy approaches have been applied for website evaluation~\cite{Pam18}. Huang \emph{et al.} \cite{huang2023improved} proposed a McDM approach based on the Technique for Order of Preference by Similarity to Ideal Solution (TOPSIS)~\cite{Chen2000} to evaluate working conditions in industrial environments, showing the importance of adapting McDM to evaluation contexts. Agrawal \emph{et al.}~\cite{Agrawal22} proposed a usability-accessibility‑based McDM approach especially aligned with WCAG 2.0 recommendations using TOPSIS and the Analytic Hierarchy Process (AHP)~\cite{Liu2020} to evaluate airline websites. }

The A/B testing technique \modif{can assess web usability since it is a} form of hypothesis testing \modif{that compares} two variants of software from the end-user perspective~\cite{Quin24}. It is widely used by marketing, communication and design professionals~\cite{Koning22}, and enables different versions of an interface to be compared in order to identify what works best~\cite{firmenich2019usability}. Thus, this technique helps to recognize the value of user feedback~\cite{king2017designing}. 

By integrating standardized tests and reports under the point of view of the user by \textit{role-playing} into A/B testing, we hypothesize that we can develop a high-quality proposal to address website usability evaluation, aiming to enhance user experience, bridge the gap between developer and user perspectives, and fostering to provide user-centered design solutions. \modif{Finally, we notice that} software engineers and interface designers could be benefited by the existence of a free software tool to assist this process.

\modif{This paper proposes an online Decision Support System (DSS) for evaluating web usability based on a linguistic McDM methodology. It can be depicted as two main components:} 
\begin{itemize}
\item \textit{The linguistic decision-making \modif{for web usability evaluation} (LDM4WUE) methodology:} is based on user perceptions and user-centered techniques such as \modif{DT}. We choose 2-tuple fuzzy linguistic model \cite{herrera2001} to handle qualitative aspects and employ fuzzy AHP \cite{chang1996applications} and TOPSIS \cite{Chen2000} to assign weights to criteria and generate usability rankings, considering users with different needs or disabilities. The methodology guides UX experts in the processes of usability evaluation incorporating user needs and expectations. This might improve the quality of the final interface design in the quest to achieve more inclusive and usable software.

\item \textit{The USE-AB-DSS:} is a DSS for A/B testing in UX engineering contexts. The tool is present in all steps of the methodology: setting up the evaluation project, collecting feedback and computing usability levels and ranking the designs according to their level of usability. USE-AB-DSS generates detailed reports very helpful when dealing role-specific view plus general reporting, providing a comprehensive and detailed view of the system usability. It integrates the processes of the LDM4WUE methodology, reducing the time of the decision-making calculus. It is a free online tool that can be used even with students of Human Computer Interaction (HCI) disciplines.

\end{itemize}

In order to showcase the practicality and utility of the proposal, a \modif{Massive Open Online Course (MOOC)} has been deployed at three different Moodle platforms used by universities of the University of Guadalajara in Mexico. The selected MOOC, the \textit{Course on Inclusive Educational Contexts: Design for All}~\cite{liliana20} (best known as DUA-MOOC), covers the teaching practices to be taken to comply with the \modif{universal design for learning} guidelines~\cite{DUA11}. We disable any assistive technologies (AT) in the three LMS although users can enable them at the operating system level (e.g. Microsoft Narrator). This way, an ideal scenario is achieved for the A/B testing comparison of the alternatives websites, since the results will shed light only on the degree of usability of each platform. 

The remaining of this paper are structured as follows. Section~\ref{sec:antecedentes} introduces the usability tests and representations to resolve a linguistic decision-making problem for website usability assessment. Section~\ref{sec:model} presents the LDM4WUE methodology, including the stages required to conduct usability tests and how to obtain a final linguistic usability score, and also, scores for each role played. Section~\ref{sec:soft} describes the implementation of the \modif{USE-AB-DSS} for comparing alternatives and managing user-generated data. Section~\ref{sec:case} presents the case study whereby all the procedures are followed in order to determine which of the three Moodle platforms offers the best usability for students and teachers. Section~\ref{sec:conclusiones} presents our conclusions and outlines future work.

\section{Preliminaries}\label{sec:antecedentes}

Our proposal is strongly related to concepts associated with website evaluation, such as web accessibility (as explained in Section~\ref{sec:accesibilidad}) and web usability (as defined in Section~\ref{sec:usabilidad}). Section~\ref{sec:2tuplemethod} presents the basic aspects of the 2-tuple linguistic representation model to explain the LDM4WUE methodology. Finally, in Section~\ref{sec:topsismethod}, we describe the TOPSIS algorithm for solving the LDM4WUE exploitation phase. 

\subsection{Web accessibility evaluation methods} \label{sec:accesibilidad}

Web accessibility is the practice of ensuring access to information on websites, especially for people with disabilities, be they visual, hearing, physical, or cognitive. The purpose of web accessibility evaluation is to measure the possibility of the site being used not only by people with disabilities but also any other person, as proposed by the \textit{design for all} paradigm.

Given that standards and recommendations for ensuring correct software accessibility have been established for years, especially in the context of websites, a number of tools now facilitate the automatic evaluation of accessibility. The World Wide Web Consortium (W3C) plays a key role in this standardization effort. Their website features a comprehensive list of 85 accessibility testing tools\footnote{W3C list of assessment tools \url{https://www.w3.org/WAI/ER/tools/}}, although only a few can effectively monitor compliance with the recently published WCAG 2.2 standard~\cite{wcag22}, checked in July 2025. The ultimate aim is to assess a website's accessibility and assign it an A, AA, or AAA label, typically displayed as an image in the site footer.

One of the most used tools for evaluating web accessibility is WAVE\footnote{Wave Accessibility testing tool \url{https://wave.webaim.org/}}, which provides a detailed report of the errors and warnings on the evaluated HTML page. \nzm{This allows developers to modify the HTML/CSS accordingly. After adjustments, the page can be reevaluated to achieve the desired accessibility level.}

\subsection{Web usability evaluation methods} \label{sec:usabilidad}

In the past, the usability of a product was connected with its user-friendliness. 
This concept is inherently subjective, making it challenging to establish standardized measurements. From a practical point of view, usability is about the experience of a user and the fact of being able to operate a system in the minimum time possible, without neglecting aesthetics and site content. Classical approaches for usability evaluation are commented below.

\begin{figure}[h]
\centering
     \includegraphics[width=12cm]{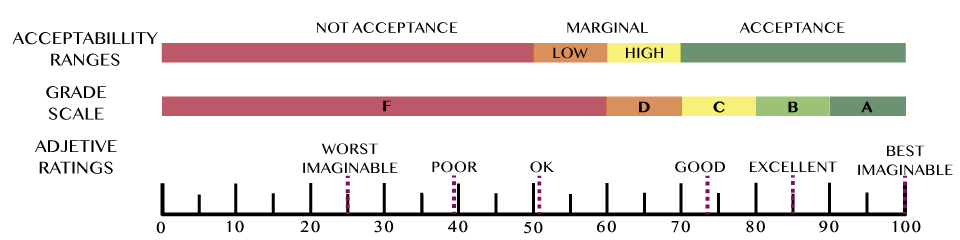}
    \caption{Several linguistic scales can be derived from the numerical SUS score. Source:~\cite{bangor2009}.}
    \label{fig:sus}
\end{figure}

\begin{description}
    \item [Inspection.] A panel of experts play an important role in measuring the usability of a system by testing the user interface. There are practical checklists such as Nielsen's heuristic~\cite{nielsen1993} or a Cognitive Walk-through for a given task flow~\cite{farzandipour2021usability}.

    \item [Inquiry.] This method focuses on data acquisition mainly by observing people during the software usage processes. The following tests fall into this category: 
    \begin{itemize}
        \item \textbf{Eye Tracking}, a solution related to neuromarketing that tracks the eyes to know the point where the gaze is fixed. This can help to better understand what attracts the customer's attention.
        \item \textbf{Ad hoc}, specifically designed tests that apply to real people with disabilities and assistive technology enabled browsers.
        \item \textbf{Focus Groups}, discussion with users and recording facilities~\cite{shahini2021usability}. 
        \item \textbf{Interviews}, usually letting the user think out loud~\cite{hussain2019interview}. 
        \item \textbf{Activity Log}s~\cite{lee2019development}, that can be further explored with data mining techniques~\cite{buenano2019use}.
    \end{itemize}
    
    \item [Usability Test.] This method collects information in real time in sessions called \textbf{usability test} (UT), with participants that are real users who will use the software. This information, therefore, has a high degree of reliability~\cite{krug2000don}. In addition, this test can detect some omissions derived from the heuristic evaluations. In order to avoid confusion, the UT must be explicitly defined by \emph{the conductor} according to the sites to be evaluated, focusing on the more frequently used tasks. Before application, it must previously be explained to \emph{the participants} (number of task, expected starting and ending time) in order to enable them to perform the test as independently as possible. This test can be conducted face-to-face or through an online session, subsequently collecting the information with tools such as Google Forms~\cite{liliana20}. 

    \item [Standardized questionnaires.] Various usability evaluation questionnaires have been developed and accredited~\cite{albert2013measuring} and these include the System Usability Scale (SUS), the Questionnaire for Users Interfaces Systems (QUIS), and the Web site Analysis and MeasureMent Inventory (WAMMI). It is also very common to use a single question for product / service satisfaction inquiry, called the Net Promoter Score (NPS)~\cite{Reichheld2004}. These tests evaluate different criteria or dimensions of usability, such as software performance, design, ease of use, user satisfaction, etc. Our model is able to linguistically incorporate the results from SUS and NPS questionnaires.

    \begin{itemize}
        \item SUS was developed by John Brooke in 1996 \cite{Brooke:1996,Brooke:2013} and is frequently used for usability evaluation in fields such as medicine ~\cite{schmettow2017extended}, mobile applications~\cite{quinn2019mobile} and services~\cite{ambarwati2021usability}. Its success is largely due to the fact that it is extremely easy to complete with only 10 Likert-scale items, it is free and available in multiple languages, and it enables the evaluation of various types of user interfaces (e.g. websites, mobile apps, TVs).  From the answers, a formula then calculates a number in the 0-100 range. However, more importantly, the SUS fits many different linguistic scales as shown in Figure~\ref{fig:sus}~\cite{bangor2009}.
        
        \item NPS~\cite{Reichheld2004,Owen2019} is used for service evaluation in companies and institutions and involves a single, simple question in order to classify customers as promoters, passive users or detractors. The user answers with a number between 0 and 10, but the NPS score is a value within the range $[-100,100]$. There is a factor of conversion from the NPS to SUS scale as can be seen in~\cite{sauro2016quantifying}.
    \end{itemize}

    \item [A/B testing.] This type of testing is used extensively in business marketing. For instance, it is used to improve performance metrics~\cite{Koning22} such as conversion and click-through rates. A/B testing compares two versions of the same product or service in an attempt to identify which version or features are considered better by the user. The importance of this technique is to recognize the value of user feedback in order to design better user experiences~\cite{king2017designing}. In the UX field, the A/B testing technique has been incorporated to compare two versions of the same site~\cite{firmenich2019usability}, for instance, to test satisfaction with dark versus light themes. \nzm{A/B testing tools include commercial platforms\footnote{VWO \url{https://vwo.com/}} and basic calculators\footnote{\url{https://www.kissmetrics.com/growth-tools/ab-significance-test/}}. However, comprehensive and free online tools for usability testing are still lacking}. Therefore, rather than only comparing Case A or Case B, there could be more alternatives. 
    
\end{description}

\subsection{A linguistic representation model for decision-making} \label{sec:2tuplemethod}
\nzm{Decision-making (DM) models are essential tools for computationally addressing complex decision scenarios in domains such as healthcare, business, and education. In the context of usability evaluation, these models enable the structured integration of expert input and user feedback to facilitate the systematic comparison of interface alternatives}. They require effective management of the expert evaluations to rank the alternatives with quality. There are multiple different representations for solving DM, among which the 2-tuple linguistic representation has a great impact~\cite{malhotra2020systematic}. It handles uncertainty through linguistic terms and prevents the loss of information by means of a continuous domain. Rooted on the fuzzy set theory, Herrera \textit{et al.} \modif{\cite{2tuplasBook}} define the functions $\Delta$ and $\Delta^{-1}$ to transform numerical values into 2-tuples and vice versa.

In our case, for usability assessments where individuals have different needs, consensus-reaching processes, which are usually embedded in group decision-making (GDM)~\cite{Yao2023}, may not be sought. The focus lies more on understanding and accommodating diverse perspectives rather than striving for uniform agreement among stakeholders.

\begin{proposition} \cite{2tuplasBook}
Let $S^{g+1} = \{s_0, \dots, s_g\}$ be a linguistic term set and $\beta \in [0,g]$ the result of an aggregation operation. The function $\Delta$ transforms a $\beta$ value to an equivalent information 2-tuple with Equation \ref{eq:2TupleDelta}. 
\begin{equation}\label{eq:2TupleDelta}
\begin{array}{l}
\Delta: [0,g] \rightarrow S^{g+1} \times [-0.5,0.5)\\
\Delta(\beta)~=~(s_i,\alpha), \;with\; 
	\left\{
	\begin{array}{l}
	    s_i \;\;\;i=round(\beta), \\
	   \alpha = \beta - i
	\end{array} 
	\right .
\end{array}
\end{equation}
where \emph{round($\cdot$)} is the usual round operation.
\end{proposition}

\begin{proposition} \cite{2tuplasBook}
The inverse function $\Delta^{-1}$ transforms a 2-tuple to its equivalent numerical value $\beta \in [0,g]$ with Equation \ref{eq:2TupleInverseDelta}.
\begin{equation}\label{eq:2TupleInverseDelta}
\Delta^{-1}(s_i,\alpha)~=~i + \alpha = \beta
\end{equation}
\end{proposition}

\subsection{A ranking method for decision-making}\label{sec:topsismethod}

The Technique for Order of Preference by Similarity to Ideal Solution, known as TOPSIS~\cite{nada16}, stands out as a multi-criteria method that facilitates the selection of the most optimal alternative. It is based on the assumption that the selected alternative is as close as possible to the positive ideal solution. \nzm{TOPSIS is selected because of its ability to rank alternatives based on proximity to ideal usability criteria, making it especially suitable for this linguistic multi-criteria decision-making context.}

Let $D$ be a normalized matrix comprising 2-tuple values $(s_{ij}, \alpha_{ij})$, where $i=1, \dots, n$ represent the evaluated alternatives and $j=1, \dots, m$ the associated criteria. We assume that the weights of the criteria are equal. TOPSIS determines the positive ideal solution $A^{+}$ and \modif{the} negative ideal solution $A^{-}$ as vectors of 2-tuples formed by:
\begin{equation}
\label{eq:a+l}  
  A^{+}= [(r^{+}_1,\alpha^{+}_1), (r^{+}_2,\alpha^{+}_2), \dots , (r^{+}_m,\alpha^{+}_m) ]
\end{equation}
\begin{equation}
\label{eq:a-l}  
  A^{-}= [(r^{-}_1,\alpha^{-}_1), (r^{-}_2,\alpha^{-}_2), \dots , (r^{-}_m,\alpha^{-}_m) ]
\end{equation}
where  
\begin{equation}
\label{eq:r+l}  
(r^{+}_j,\alpha^{+}_j) = \left\{ \underset{i}{max} \{ (s_{ij},\alpha_{ij})\} \right\}, j=1, \dots, m
\end{equation}
\begin{equation}
\label{eq:r-l}  
(r^{-}_j,\alpha^{-}_j) = \left\{ \underset{i}{min} \{ (s_{ij},\alpha_{ij})\} \right\}, j=1, \dots, m
\end{equation}

The separation measures $D^{+}_i$ and $D^{-}_i$ of each alternative from the positive ideal solution and the negative ideal solution are then computed based on the Euclidean distance:
\begin{equation}
\label{eq:d+l}    
D^{+}_i = \Delta \sqrt{\sum_{j=1}^m \left(\Delta^{-1} (s_{ij}, \alpha_{ij})-\Delta^{-1}(r^{+}_j,\alpha^{+}_j)\right)^2}
\end{equation}

\begin{equation}
\label{eq:d-l}
D^{-}_i = \Delta \sqrt{\sum_{j=1}^m \left(\Delta^{-1} (s_{ij}, \alpha_{ij})-\Delta^{-1}(r^{-}_j,\alpha^{-}_j)\right)^2}
\end{equation}

The coefficient of relative closeness to the ideal solutions of each alternative \modif{$i=1, \dots, n$} in relation to the positive ideal solution $A^{+}$ is then calculated:
\begin{equation}
\label{eq:rc+l}
RC_i = \Delta \left(\frac{\Delta^{-1}(D^{-}_i)}{\Delta^{-1}(D^{+}_i) + \Delta^{-1}(D^{-}_i)} \right)
\end{equation}
where $0 \leq \Delta^{-1}(RC_i) \leq 1$.

\section{A Multi-expert Multi-criteria Linguistic Decision-Making for Web Usability Evaluation Methodology}
\label{sec:model}

This section explains the multi-expert multi-criteria linguistic \modif{DM for web usability evaluation methodology,} \textit{i.e.}, the \textit{LDM4WUE} methodology. It merges various sources of information, incorporates the use of standardized tests and takes into account the approximation to human reasoning by means of the linguistic transformation of user judgments. The methodology resembles A/B testing, which compares two versions of the same product, with the advantage of preserving its usability even when assessing more than two alternatives. Section~\ref{sec:high} highlights the benefits of the application of the methodology in the context of user interface design. Section~\ref{sec:DM} describes the phases for solving the underlying linguistic decision-making problem. In order to document each phase, the problem definition is explained in Section~\ref{problemstate} and the \textit{role-playing} technique is explained in Section~\ref{sec:roles}. This is followed by the data eliciting and gathering phase in Section~\ref{sub:EIU}, the aggregation phase in Section~\ref{sub:agregacion} and the exploitation phase in Section~\ref{sub:ranking}.

\subsection{Main characteristics of the LDM4WUE methodology} \label{sec:high}
\begin{figure}[htb]
\centering
 \includegraphics[width=0.4\textwidth]{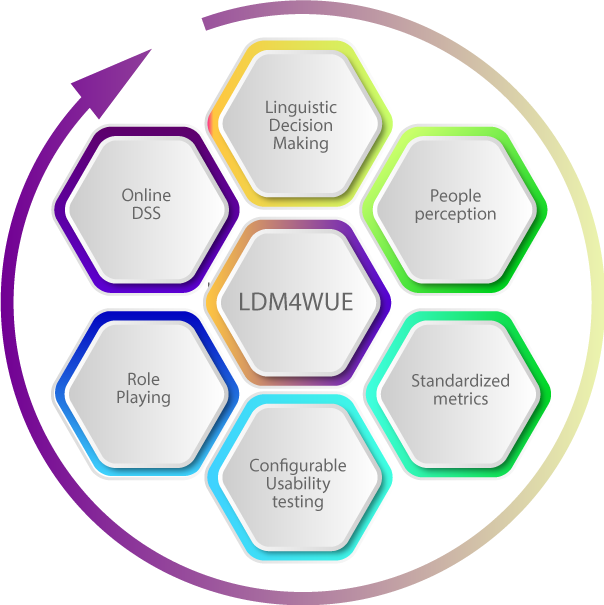}
 \caption{Key elements of the user-centered LDM4WUE methodology.}
\label{fig:hexa}
\end{figure}

Web usability evaluation, from a user-centered \modif{perspective,} should emphasize satisfaction with the use of the site and data provided by the UX experts or end-users. \modif{The LDM4WUE methodology is designed to enhance this perspective by combining linguistic decision-making with standardized usability assessments and a highly configurable evaluation pipeline. Its key components are shown in Figure~\ref{fig:hexa} and are described below:}
\begin{itemize}
    \item \modif{\textbf{LDM4WUE applies linguistic decision making.} Evaluations are provided through linguistic terms instead of numerical values to align with the nature of human judgments. It enables more interpretable inputs and outputs for the decision-making process.}

    \item \textbf{LDM4WUE \modif{considers} people's perceptions.} The \modif{evaluations} are obtained from two groups of users (UX experts and website end-users) and managed with the 2-tuple linguistic computational model~\cite{herrera2001}. 
    It acknowledges varying expertise levels, perspectives, and contributions, enabling more informed and balanced decision-making processes.

    \item \textbf{LDM4WUE uses \modif{standardize tests.}} It combines custom and standardized usability tests with accessibility tests from a linguistic perspective. Results are reported in the same domain of significance, qualifying the website's usability with the adjective SUS scale~\cite{bangor2009}.
    
    \item \textbf{LDM4WUE is configurable}. It enables the application of as many tests as necessary. Additionally, it can be configured with personalized usability tests thanks to the incorporation of the \emph{usability testing} concept, for which we can define the most appropriate number of tasks and find the most suitable task-estimation time. 

    \item \textbf{LDM4WUE \modif{incorporates \textit{role-playing}}.} It relies on the design thinking paradigm \modif{to simulate user diversity} by using techniques such as \textit{role-playing}. Evaluating a website from the perspective of specific roles (e.g. visually impaired, elderly, stressed) helps to observe compliance with the \textit{design for all} principle. 
    
    \item \modif{\textbf{LDM4WUE can be implemented as a DSS.} To facilitate the methodology application, we integrate it in a free, online DSS (named \textit{USE-AB-DSS}) that supports A/B testing, data collection, ranking generation, and automated reporting.}
 \end{itemize} 
 
\subsection{Flowchart of the LDM4WUE methodology} \label{sec:DM}

The LDM4WUE methodology \modif{follows} the standard decision-making problem solving steps~\cite{Herrera2000} with \modif{particular} adaptations to determine the linguistic variable of usability. The proposed solving schema for web usability evaluation outlines \modif{five} stages:

\begin{enumerate}
    \item \textbf{Problem description.} We establish a moderator who defines the set of alternatives as websites to compare, their associated criteria as tests for web usability assessment and the set of users that evaluate the alternative websites based on the criteria.

    \item \textbf{Empathy and role-playing.} This phase consists in explaining the objectives of the usability evaluation and the role-playing technique. \modif{A moderator} defines a set of roles that users can play when evaluating. The \modif{moderator} allows the participants extra time so that they can choose a role \modif{or uses a} die to make this point more dynamic.
    
    \item \textbf{Elicitation of user information.} The users play particular roles and individually evaluate the alternatives based on the relevant criteria (in this case, four selected tests). We gather the evaluations of each test and then build an individual decision matrix for each user playing a role. The evaluations are provided in different formats depending on the test. We computationally integrate the evaluations of each test to obtain linguistic terms. We consider $S^g$=$\{s_0, \dots, s_{g-1}\}$ as a linguistic term set of $g$ linguistic term elements \modif{and handle} 2-tuple linguistic term representations.
    
    \item \textbf{Collective aggregation.} This phase aggregates the individual user evaluations to obtain the collective aggregation. The linguistic term evaluations of the criteria tests belong to different linguistic term sets that are represented as a hierarchy. In order to aggregate \modif{them}, we first unify the various linguistic terms so that they all belong to the deepest level of the hierarchy, which in our case is the term set $S^9$. We then aggregate the individual evaluations by roles to obtain a unified collective decision matrix for each role. Finally, we aggregate the previous matrices into the unified collective decision that compiles every user evaluation into a single matrix. The aggregations are computed by the 2-tuple weighted average (\textit{2TWA}) operator. 
    
    \item \textbf{Exploitation.} In order to rank the alternatives from the best to the worst assessed, we apply the TOPSIS algorithm~\cite{Chen2000,nada16}. For convenience, in addition to having a ranking, it is possible to present linguistic output information on a specific scale. In our case, we perform this step before generating the report using the adjective SUS scale.
\end{enumerate}

\begin{figure}[htb]
\centering
 \includegraphics[width=0.98\textwidth]{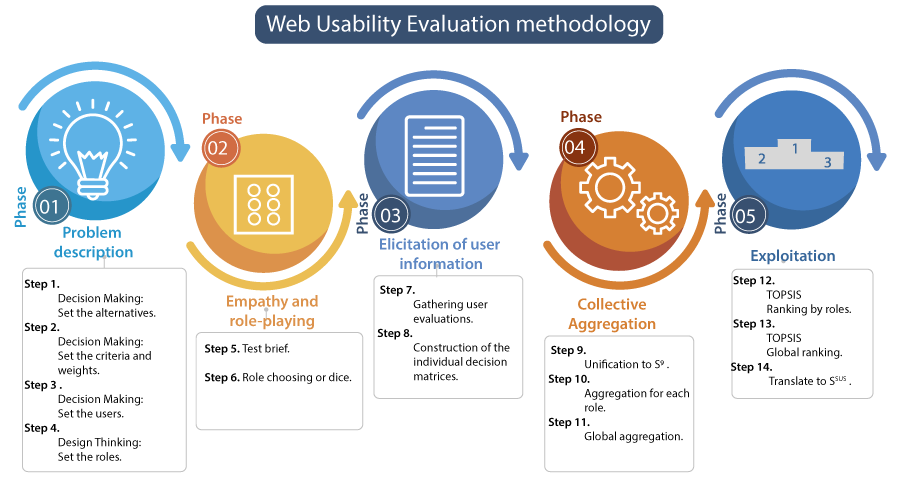}
 \caption{Flowchart of the multi-expert multi-criteria linguistic decision-making for the web usability evaluation methodology.}
\label{fig:model}
\end{figure}

Figure \ref{fig:model} illustrates the \modif{five stages, which are depicted in twelve steps, of the LDM4WUE methodology}. Further details about each step are provided in the following sections.

\subsection{Phase 1. Problem description} \label{problemstate}
We establish a \textit{moderator} (usually the site developer or the interface designer) who helps to set the following elements: alternatives, which are the websites or online platforms to evaluate; criteria, which are accessibility instruments and usability tests; decision makers or users, consisting of both end-users and experts who evaluate the alternatives based on the criteria; roles, the moderator could select ---roles regarding to touch, hear, see, or speak conditions---; and weights, that can be criteria weights, user weights and role weights.

\bigskip
\label{sub:alternatives}\textbf{Step 1. Definition of the alternative set}. We define a set of alternatives $A = \{A_1,\dots,A_n\}$ as websites, website versions, web pages, or online platforms that have similar objectives. We refer to each alternative by $A_i, (i=1,\dots,n)$.

\bigskip
\label{sub:criteria} \textbf{Step 2. Definition of the criteria set}. We define a set of criteria $C=\{C_1, \dots, C_m\}$ to be evaluated \modif{and} each one is referenced by $C_j, (j=1,\dots,m)$. With our software, the moderator can create an A/B testing with $m$ or a subset of tests. \modif{In this proposal we set four tests, setting $m=4$:}
    \begin{itemize}
        \item $C_1$ $\cong$ the SUS questionnaire comprises 10 Likert-scale items.
        \item $C_2$ $\cong$ the NPS is completed with a single number between 0 and 10. 
        \item $C_3$ $\cong$ the UT is completed \modif{answering multiple tasks which are filled} with a boolean \textit{task-done}, the \textit{task-time} in seconds, and a linguistic term to express task satisfaction. This term is selected from $S^5= \{$Unsatisfied, Dissatisfied, Indifferent, Satisfied, Very satisfied$\}$.
        \item $C_4$ $\cong$ the ACC is completed with the Accessibility report value. This term is selected from $S^3= \{A, AA, AAA\}$. 
    \end{itemize}

\modif{We consider criteria weights since some criteria may have different importance depending on the requirements of the problem. We denote the vector of criteria weights by $WC' = \{WC'_1, \dots, WC'_m\}$.} This vector is normalized by generating the criteria weights normalized vector $WC = \{WC_1, \dots, WC_m\}$ which verifies $\sum_{j=1}^{m}WC_{j}=1$. In order to precisely determine the criteria weights, we suggest the application of the fuzzy extended AHP (FAHP)~\cite{chang1996applications} method. \modif{It} facilitates organizing criteria into a hierarchy and enables the detection of inconsistencies through a consistency analysis between judgments. \modif{The following steps describe how the criteria weights are obtained.}

\begin{itemize}
    \item \textbf{Step 2.1. Obtain pairwise judgments regarding the importance of criteria.} The moderator provides pairwise judgments so that we can derive the priority scale for each criterion. We complete \modif{a criteria preferences ($CP$) matrix} with triangular fuzzy numbers (TFN)~\cite{dubois1980}. 
    
    \modif{Particularly,} the moderator \modif{compare criteria by means of} the set $S_{CP}^5=\{$\textit{Equally important}, \textit{Moderately important}, \textit{Very important}, \textit{Strongly important}, \textit{Absolute}$\}$, where \modif{the semantic of the linguistic terms is represented by TFNs denoted by $(low, me\-dium, upper) = (l, m, u)$} ~\cite{chang1996applications}. The $m\times m$ \modif{$CP$ matrix} is created \modif{collecting the fuzzy numbers associated to the linguistic terms} provided by the moderator \modif{as shown in Equation \ref{TFN}: }
    \begin{equation} \label{TFN}
    CP_{j,j'} = (l_{j,j'}, m_{j,j'}, u_{j,j'}), \forall j \leq j'; \; \; j,j'=1, \dots, m 
    \end{equation}
    and their values are completed using Equation~\ref{completarCP}: 
    \begin{equation}  \label{completarCP}
    CP_{j',j} = (l_{j',j}, m_{j',j}, u_{j',j}) = \left(\frac{1}{u_{j',j}}, \frac{1}{m_{j',j}}, \frac{1}{l_{j',j}}\right), \forall j > j'.    
    \end{equation}

    \item \textbf{Step 2.2. Compute the fuzzy synthetic extension}. The fuzzy synthetic extension for criterion $C_j, \modif{(j=1, \dots, m)}$ is calculated using Equations~\ref{eq:extsintetica1} -~\ref{eq:extsintetica3}:
    \begin{equation}   \label{eq:extsintetica1}
    s_{j} = (l_{j}, m_{j}, u_{j}) = \sum_{j'=1}^{m}{M_{C_j}^{j'}} \otimes [\sum_{j=1}^{m}{\sum_{j'=1}^{m}{M_{C_j}^{j'}}}]^{-1}, \forall j=1, \dots, m
    \end{equation}

    where  
    \begin{equation} \label{eq:extsintetica2}
    \sum_{j'=1}^{m}M_{C_j}^{j'} = \left(\sum_{j'=1}^{m}l_{j,j'}, \sum_{j'=1}^{m}m_{j,j'}, \sum_{j'=1}^{m}u_{j,j'}\right), \forall j=1, \dots, m, 
    \end{equation}

    \begin{equation}\label{eq:extsintetica3}
    [\sum_{j=1}^{m}{\sum_{j'=1}^{m}{M_{C_j}^{j'}}}]^{-1} = \left(\frac{1}{\sum_{j=1}^{m}\sum_{j'=1}^{m}u_{j,j'}}, \frac{1}{\sum_{j'=1}^{m}\sum_{j'=1}^{m}m_{j,j'}}, \frac{1}{\sum_{j=1}^{m}\sum_{j'=1}^{m}l_{j,j'}}\right).
    \end{equation}

    \item \textbf{Step 2.3. Possibility Degree}. We obtain the degrees of possibility of the elements $s_j$ and $s_{j'}$, $\forall j, j'= 1, \dots, m$, $j \neq j'$ using Equation~\ref{eq:degreeofpossibility}:
    \begin{equation} \label{eq:degreeofpossibility}
    V(s_j \geq s_{j'})=
    \begin{cases}
        1,\quad \quad \quad \quad \quad \quad \quad \quad \quad if \; s_j \geq s_{j'}\\
        0,\quad \quad \quad \quad \quad \quad \quad \quad \quad if \; l_{j'} \geq u_{j}\\
        \dfrac{l_{j'} - u_{j}}{(m_{j} - u_{j}) - m_{j'} - l_{j'}}, \; in \; any \; other \; case
    &\end{cases}
    \end{equation}

    \item \textbf{Step 2.4. Obtaining the vectors of \modif{criteria weights}}. Finally, we compute the weight of each criterion $C_j, (j = 1, \dots, m)$ using Equation~\ref{eq:weight}: 
    \begin{equation}\label{eq:weight}
    WC'_j = min[V(s_j \geq s_{j'})],  \forall j, j'= 1, \dots, m, j \neq j',
    \end{equation}
    
    and these are normalized to obtain the final weight for each criterion $C_j, (j = 1, \dots, m)$ by applying Equation~\ref{eq:pesonorm}: 
    \begin{equation}\label{eq:pesonorm}
    WC_j = \dfrac{WC'_j}{\sum_{j=1}^{m}{WC'_j}}.
    \end{equation}

\end{itemize}

\bigskip
\label{sub:users}\textbf{Step 3. Definition of the user set}. Two groups of users are considered: experts and end-users. Let $E = \{E_1,\dots,E_p\}$ be a set of experts with knowledge in some area of technology, interfaces, or user experience, where $p$ is the total number of experts. Let $D = \{D_1,\dots,D_q\}$ be a set of end-users, where $q$ is the total number of non-expert users. The set of users $U = E \cup D$ is the union of experts and end-users, and each user is referenced by $U_k, (k = 1,\dots,u)$ where $u = p+q$. \modif{A weight} is associated with each user group: $WE \in [0,1]$ in the case of experts \modif{and} $WD \in [0,1]$ for end-users. Both values are set directly by the moderator. The \modif{vector of user weights} $WU = \{WU_1, \dots, WU_u\}$ is completed with values according to whether \modif{user} belongs to one of the two groups. For example, if we have one expert and two end-users, this vector is $WU = \{ WE, WD, WD\}$. Although we do not require $WE+WD=1$, we \modif{will require to compute diverse normalizations} of $WU$ as it is described in following steps.

\bigskip
\label{sub:roles}\textbf{Step 4. Definition of the set of roles}. Since the LDM4WUE methodology aims to focus on end-users, it \modif{then relies} on design thinking with the role-playing technique (see Section \ref{sec:roles}) to capture the end-user's needs. Let $R = \{R_1,\dots,R_r\}$ be the set of roles determined by the moderator where the possible roles are $R_l, (l = 1,\dots,r)$. Each user plays at least one role in which they evaluate all the alternatives, and the importance of each role varies according to the requirements of the problem. 
Let $WR' = \{WR'_1,\dots, WR'_r\}$ be the vector of weights associated to the roles set directly \modif{set} by the moderator. This vector is normalized by obtaining the vector of role-playing weights $WR = \{WR_1, \dots, WR_r\}$ that verifies $\sum_{l=1}^{r}WR_{l}=1$.

\subsection{Phase 2. Empathy and role-playing} \label{sec:roles}

Design Thinking (DT) can be defined as ``a human-centered innovation process that emphasizes observation, collaboration, rapid learning, visualization of ideas, rapid prototyping, and simultaneous business analysis'' \cite{lockwood2010design}. It is fully capable of understanding people's needs by establishing well-defined phases and by applying a number of  tools to conceptually analyze user needs and identify software development requirements. Given its user-centered nature, this type of methodology focuses on the collection of user characteristics and needs. 

\begin{figure}[htb]  
\centering
 \includegraphics[width=0.98\textwidth]{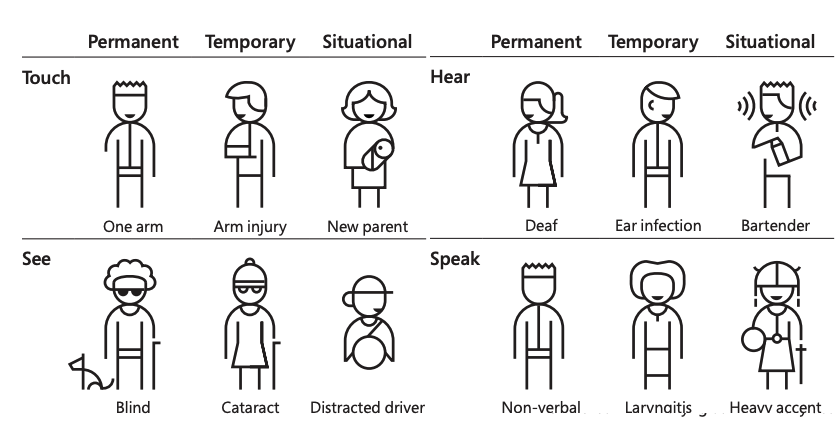}
 \caption{The decision makers can either express their opinions about their own situations or can role\-play to empathize with real user needs. Source:~\cite{MS2016}.}
\label{fig:roles}
\end{figure}

We apply \textit{role-playing} in our \modif{methodology} to allow people to express some temporary situation in their lives such as a broken arm or even their mood. By linking a role to the assessment, the UX expert who conducts the A/B testing can also apply several assessments for each of the defined \textit{personas} (an archetype of a user that helps designers and developers empathize with people with special needs~\cite{bookUX12}). For instance, they can play the role of a foreign student visiting the university website or empathize with a visually-impaired person. Numerous possibilities exist for defining these roles, as illustrated by the examples in Figure~\ref{fig:roles}~\cite{MS2016}.

\bigskip
\textbf{Step 5. Test briefing}. There are several approaches for measuring the usability of a system (as we previously described in Section \ref{sec:usabilidad}), and one of the best ways is to observe how users perform on the simplest or the most complex tasks. This is achieved with a \modif{UT}. These are scripts\footnote{Resources of `Don't make me think' book \url{https://sensible.com/download-files/}} from the book `Don't make me think'~\cite{krug2000don} that help to conduct a usability test such as instructions for the person who's going to observe the users, or a list of neutral things that the \modif{moderator} can say while the participant is performing the tasks. However, the best way of guaranteeing UT success is to brief the participants in advance to let them know what they are going to do.

\bigskip
\textbf{Step 6. Role choosing or dice}. In order to empathize with people and multiple needs, we use \textit{role-playing} as a technique \modif{that allows} to simulate varied conditions. One way of gamifying this step is to use dice. Let us suppose that we throw three dice: the first die determines the age (adolescent, elderly, etc.), the second a particular physical condition (sight, hearing, etc.), and the third a mood (depressed, stressed, tired, etc.). We could therefore role-play\modif{, for example,} a motivated elderly man with impaired vision. 

\subsection{Phase 3. Elicitation of user information}
\label{sub:EIU}

Every user evaluates criteria $C_1,C_2,C_3$ (common for $U$) and $C_4$ (only for $E$) in a particular way according to the test to which \modif{the criterion} refers and each single assessment is attached to the role selected. Each user $U_k$, playing a role $R_l$, assesses each alternative $A_i$ and according to criteria $C_j$. Generally speaking, this configures the \textit{individual decision} matrix $ID^{k,l} = (ID^{k,l}_{ij})_{n \times m}$. We will now proceed to describe \modif{the steps that} explain in detail how to construct the $ID$ matrices for each \modif{test} or criterion.

\bigskip
\label{sec:Criterias}\textbf{Step 7. Gathering user evaluations}. The user $U_k$ playing the role $R_l$ evaluates an alternative $A_i$ by performing one or more tests $C$ to qualify the usability of $A_i$. The information provided by each possible test is shown below:

\begin{description}

\item [$C_1 \cong$ System Usability Scale (SUS).] The user answers ten questions following a Likert scale with five response options for each one. Odd questions have a positive connotation while even questions have a negative tone. For each alternative $A_i$, we denote the ten responses provided by the user $U_k$ playing the role $R_l$ as $x^{k,l,i}_h, h = 1, \dots , 10$. We obtain the SUS score of user $U_k$ playing the role $R_l$ for each alternative $A_i, (i = 1, \dots , n)$, by:

\begin{equation}
    \label{eq:sus}SUS\_score^{k,l}_{i} = 2.5 \times \sum_{h=1}^{5}[(x_{2h-1}^{k,l,i} -1)+(5-x_{2h}^{k,l,i})], \forall i=1, \dots, n
\end{equation} 

For each user $U_k$ playing the role $R_l$, the value $SUS\_score^{k,l}_{i} \in [0, 100]$ is available for each alternative $A_i, (i = 1, \dots , n)$ \modif{that enables to evaluate the criterion $C_1$}. 

\item [$C_2 \cong$ Net Promoter Score (NPS).] The user faces a single question known as Likelihood to Recommend (LTR): \textit{How likely are you to recommend the website represented by $A_{i}$?}. The answer must be an integer value belonging to the interval $[0,10]$ so that values close to 0 mean that you would not recommend the website while values closer to 10 mean that you would recommend it. Thus, the user's LTR score $U_k$ is obtained by playing the role $R_l$ for each alternative $A_i, (i=1, \dots, n)$ ($NPS\_LTR\_score^{k,l}_{i}$) directly through the answer to the previous question. Users who answer with the values 10 or 9 are known as promoters, those who answer with 8 or 7 are called passive, and if they answer with another value, they are known as detractors. In a complementary way, the NPS value associated with the evaluation of all users can be obtained as the percentage of promoters minus the percentage of detractors. This calculation is frequently performed with online tools.\footnote{NPS calculator \url{https://npscalculator.com/}}

In summary, for each user $U_k$ playing the role $R_l$, there is a value $NPS\_LTR\_score^{k,l}_{i} \in [0, 10]$ for each alternative $A_i, (i=1, \dots, n)$ that enables to evaluate the criterion $C_2$.

\item [$C_3 \cong $ Usability Test (UT).] The end-user must answer as many questions as the moderator poses by setting a UT of $d$ tasks to be performed. These $d$ tasks define our usability test $UT=\{q_1, \dots, q_d\}$. The user $U_k$'s responses playing the role $R_l$ for alternative $A_i, (i=1, \dots, n)$ to each task $q_v, v=1,\dots,d$ are the following three measures:
 \begin{itemize}
    \item Efficiency ($Efficiency\_score^{k,l}_{i}(q_v)$). It establishes whether the user has managed to perform the requested task in an adequate amount of time. The user tracks the time taken ($time^{k,l}_{i}(q_v)$) and our system compares it with the moderator's estimate for maximum time ($MaxTime(q_v)$). This measure can take two values: 1 if the user completed $q_v$ under the estimated time, \textit{i.e.} if $time^{k,l}_{i}(q_v) \leq MaxTime(q_v)$, and 0 \modif{in other case}.
    \item Success ($Success\_score^{k,l}_{i}(q_v)$). 
    The user indicates whether \modif{he/she was} successful or unsuccessful in performing the task. This measure can take two values: 1 if successful and 0 if unsuccessful.
    \item Satisfaction ($Satisfaction\_score^{k,l}_{i}(q_v)$). The user indicates the feeling \modif{that} experienced while solving the task. This is expressed by one adjective out of five possible ones: \textit{unsatisfied}, \textit{dissatisfied}, \textit{indifferent}, \textit{satisfied}, and \textit{very satisfied} which correspond to the five linguistic terms of $S^5 = \{s_0, s_1, s_2, s_3, s_4\}$. 
\end{itemize}

We compute the success, efficiency, and satisfaction measures for each question based on each user's responses in order to discover the tasks that are remarkably complex for them. Furthermore, we calculate these three metrics for each possible role played to identify if there are tasks that are more difficult for a specific type of user profile. From $WU'$, we \modif{have to generate} $r$ \modif{normalized} weighted vectors $WU^{l}$ with the normalized weights of all the users playing the role $R_l, (l=1, \dots, r$). We then compute the success, efficiency, and satisfaction associated to the task $q_v, (v=1, \dots, d)$ and the given role $R_l$ using the \modif{Equations \ref{eq:efficiency}, \ref{eq:success} and \ref{eq:satisfaction}}:

\begin{equation}
\label{eq:efficiency}
Efficiency^{l}_{i}(q_v) = 100 \times \dfrac{\sum_{k=1}^{u^{l}}Efficiency\_score^{k,l}_{i}(q_v)}{u^{l}}
\end{equation}

\begin{equation}
\label{eq:success}
Success^{l}_{i}(q_v) = 100 \times \dfrac{\sum_{k=1}^{u^{l}}Success\_score^{k,l}_{i}(q_v)}{u^{l}}
\end{equation}

\begin{equation}
\label{eq:satisfaction}
Satisfaction^{l}_{i}(q_v) = \Delta\left(\sum_{k=1}^{u^{l}}\Delta^{-1}(Satisfaction\_score^{k,l}_{i}(q_v),0)\times WU^{l}_{k}\right)
\end{equation}
where $u^{l}$ is the number of users playing the role $R_l$. 

The success and efficiency metrics are percentages \modif{(which are then transformed to the unit interval)} while the satisfaction metric is a 2-tuple linguistic value. All this information is collected in the usability report that is complementary to the ranking solutions. 

For each user $U_k$ playing the role $R_l$ and for each task $q_v$, three values are then available: $Success\_score^{k,l}_{i}(q_v) \in [0, 1]$, $Efficiency\_score^{k,l}_{i}(q_v) \in [0, 1]$, and $Satisfaction\_score^{k,l}_{i}(q_v) \in S^5$. \modif{The satisfaction metric (success and efficiency are just secondary metrics for the report) enables to evaluate the criterion $C_3$.}

\item [$C_4 \cong $ Accessibility (ACC).] This test should only be performed by expert users. First, experts test alternatives using the WAVE online tool that lists errors and warnings in possible areas for improvement of the website or alternative. Given this report, the expert user $U_k$ playing the role of $R_l$ assigns a rating ($Acc\_score^{k,l}_{i}$) for each alternative $A_i, (i=1, \dots, n)$. This score measures the accessibility of the alternative, so it corresponds to the \textit{A}, \textit{AA}, and \textit{AAA} conformance criteria with current web content accessibility guidelines \cite{wcag22}. The \textit{A} label indicates the lowest level, while the \textit{AAA} label indicates the highest level and therefore the highest quality.

In conclusion, for each expert user $U_k$ playing the role $R_l$, a value $Acc\_{score}^{k,l}_{i} \in S^3 = \{A, AA, AAA\}$ is available for each alternative $A_i, i=1, \dots, n$ \modif{that enables to evaluate the criterion $C_4$.} 

\end{description}

\bigskip
\label{sub:construccion}\textbf{Step 8. Construction of the individual decision matrices}. The results of the \modif{four} tests performed by the users \modif{provide} very varied information. We consider linguistic models in order to interpret all the information together. For this purpose, this phase builds a matrix for each user playing a role that collects the results of all the performed tests, \textit{i.e.}, the evaluations they provide, represented by a linguistic approach. For each user $U_k$ playing the role $R_l$, the individual decision matrix $ID^{k,l} = (ID_{ij}^{k,l})_{n\times m}$ is constructed. Its elements correspond to the information provided by the user evaluations according to each criterion $C_j, j=1, \dots, m$ as follows:

\begin{description}
    
\item [$C_1 \cong SUS$.] The $SUS\_{score}^{k,l}_{i}, (i=1, \dots, n$) scores of user $U_k$ playing the role $R_l$ are available. These scores are values in the interval [0, 100]. If that user with that role does not evaluate alternative $A_i$ for this criterion, $SUS\_{score}^{k,l}_{i} = \{\emptyset\}$.

Bangor~\cite{bangor2009} proposes an adjective-based ranking within the SUS scale. More specifically, Figure~\ref{fig:sus} shows the SUS score equivalence in the interval $[0,100]$ in a set of unbalanced terms. We will now explain how this equivalence is performed. We define the \textit{adjective SUS} as an unbalanced set of 7 linguistic terms $\{s_0^{sus}, s_1^{sus},s_2^{sus}$, $s_3^{sus}, s_4^{sus}, s_5^{sus}, s_6^{sus}\}$ which is referred to as ${S_{SUS}} = \{$\textit{None},  \textit{Worst Imaginable},  \textit{Poor}, \textit{Ok}, \textit{Good}, \textit{Excellent}, \textit{Best Imaginable}$\}$ = $\{N,WI,P,O,G,E,BI\}$. In order to manage this term set, we jointly consider the 2-tuple representation and the hierarchical linguistic structures. To begin with, we need to choose a suitable hierarchical linguistic structure and assign the associated semantics to each term using the different levels of the hierarchy. The $S_{SUS}$ scale is therefore constructed by means of two steps:

    \begin{enumerate}

        \item Define a linguistic hierarchy $LH = \cup_{t}l(t,n(t))$. Each level of the hierarchy represents $S^{n(t)}$ and is denoted as $l(t,n(t))$, where $t$ denotes the level of the hierarchy and $n(t)$ express the granularity of the set of terms in that level, \textit{i.e.}, the number of elements available to it. We set level 1 as $l(1,3)$ to partition the scale from the center and generate the next level as $l(t+1,2n(t) - 1)$. Therefore, the second level of the hierarchy is $l(2,5)$. We set a third level $l(3,9)$ in order to adapt every $S_{SUS}$ term. In other words, at level $t=1$ of the hierarchy, we have $n(1) =3$, at level $t=2$, we have $n(2) = 5$, and at level $t=3$, we have $n(3) =9$. This enables us to establish the hierarchy $LH = S^3 \cup S^5 \cup S^9$. 
        
        \item Represent the unbalanced terms of $S_{SUS}$ in $LH$. If we apply the procedure raised in \cite{herrera2008fuzzy}, we find that $S_{SUS}$ is represented in $LH$ using the linguistic labels of levels 2 and 3 of the hierarchy as shown in Figure~\ref{fig:levels_sus}. The linguistic terms of $S_{SUS}$ belong to different levels of the $LH$ hierarchy. The semantic representation of those terms is shown in Figure \ref{fig:labels_sus} and corresponds to $N \leftarrow s_0^5$; $WI \leftarrow \overline{s_1^5}\cup \underline{s_2^9}$; $P \leftarrow s_3^9$; $OK \leftarrow \overline{s_4^9}\cup \underline{s_2^5}$; $G \leftarrow \overline{s_3^5}\cup \underline{s_6^9}$; $E \leftarrow s_7^9$; $BI \leftarrow s_8^9$. 
        
        \begin{figure}[h]
            \centering
            \includegraphics[width=10cm]{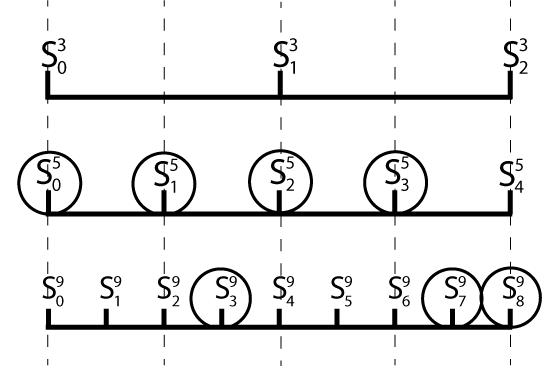}
            \caption{$S_{SUS}$ elements are defined according to the hierarchy $LH = S^3 \cup S^5 \cup S^9$ and marked with circles.}
            \label{fig:levels_sus}
        \end{figure}

        \begin{figure}[h]
            \centering
            \includegraphics[width=11cm]{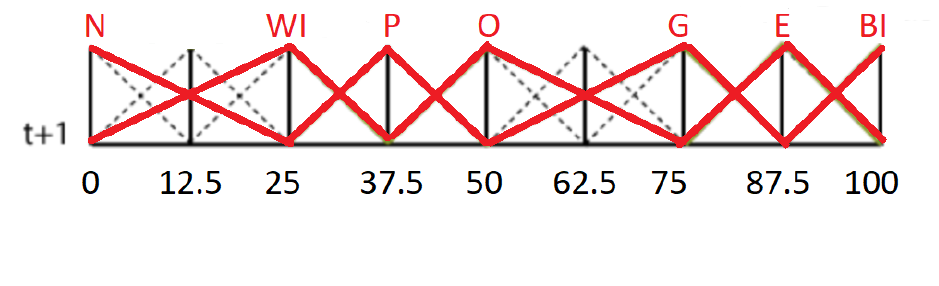}
            \caption{Semantic representation of the $S_{SUS}$ scale.}
            \label{fig:labels_sus}
        \end{figure}

        In order to work comfortably with this information, we define the set of term components $(TC)$ associated with each linguistic term of $S_{SUS}$ as the set of terms belonging to some level of the hierarchy that make up the semantics of the term in question as follows: 
        
        \begin{itemize}\label{sec:TC}
            \item $TC(s_0^{sus}) = \{s_0^5\}$
            \item $TC(s_1^{sus}) = \{s_1^5, s_2^9\}$
            \item $TC(s_2^{sus}) = \{s_3^9\}$
            \item $TC(s_3^{sus}) = \{s_4^9, s_2^5\}$
            \item $TC(s_4^{sus}) = \{s_3^5, s_6^9\}$
            \item $TC(s_5^{sus}) = \{s_7^9\}$
            \item $TC(s_6^{sus}) = \{s_8^9\}$
        \end{itemize}
        \end{enumerate}

        Thus far, we have established the hierarchical structure associated with $S_{SUS}$ as well as the semantic associated with each term. \modif{Next, we} transform the $SUS\_{score}^{k,l}_{i}, i=1, \dots, n$ scores to the $S_{SUS}$ scale. \modif{To do that}, Definition \ref{def:SUS} presents the function that enables SUS score values belonging to the interval $[0,100]$ to be transformed to the $S_{SUS}$ scale:
    
        \begin{definition}
            \label{def:SUS}
            Let $Score \in [0,100]$ be the score obtained after taking a SUS test (typically online calculator are available\footnote{SUS calculator \url{https://uiuxtrend.com/sus-calculator/}}). Let be $S_{SUS} = \{s_i^{sus}; i=0,\dots,6\}$ the unbalanced SUS linguistic scale. Let $LH = S^3 \cup S^5 \cup S^9$ be the linguistic hierarchy associated with $S_{SUS}$ such that $S'^{n(t)} = \{s_0^{n(t)}, \dots, s_{n(t)-1}^{n(t)}\}$, $t=1,2,3$. We obtain the 2-tuple belonging to the $t$ level of $LH$ associated with the $Score$ as: 
           
            \begin{equation}
            \begin{split}
                (s', \alpha') = \Delta \left( \frac{(n(t)-1) Score}{100} \right) 
            \end{split}
            \end{equation}

            where $t=2$ if $Score \in [0, 25] \cup [50, 75]$ and $t=3$ if $Score \in (25, 50) \cup (75, 100]$.

            We define a transformation function for SUS ($TF_{SUS}$) that associates the SUS score with its respective unbalanced linguistic 2-tuple as:
            \begin{equation}
            \begin{split}
                TF_{SUS}:& [0,100] \rightarrow (S_{SUS}\times[-0.5, 0.5))\\
                TF_{SUS}&(Score) = (s_i^{sus}, \alpha') \; \; | \; \; s' \in TC(s_i^{sus})  
            \end{split}
            \end{equation}
        \end{definition}

        \begin{example}
        Let $Score=53$ be the calculated value for $C_1$. This gives $t=2$ (since $Score \in [50,75]$) and, therefore, the 2-tuple associated with level $2$ of the hierarchy is $\Delta \left( \frac{4\times 53}{100} \right) = (s_2^5, 0.12)$. The associated unbalanced linguistic 2-tuple is then $TF_{SUS}(53)=(s_3^{sus},0.12)$ which is interpreted as usability=$(OK,0.12)$.
        \end{example}

        We transform the values $SUS\_score^{k,l}_{i}, i=1, \dots, n$ given by user $U_k$ playing the role $R_l$, into linguistic 2-tuples associated with the scale $S_{SUS}$ by applying Definition \ref{def:SUS} and compile them into the $ID^{k,l}$ matrices. More specifically, for each user $U_k$ playing the role $R_l$, we complete the first column ($j=1$) of its associated matrix $ID^{k,l}$ as follows:
        \begin{equation}
        \label{eq:TFSus}
        ID^{k,l}_{i1} = TF_{SUS}(SUS\_score^{k,l}_{i}) \; \; \forall i=1, \dots, n
        \end{equation}
    
 \item [$C2 \cong NPS$.] NPS values $NPS\_LTR\_score^{k,l}_{i}$ of user $U_k$ playing the role $R_{l}$ are available. These scores are values from 0 to 10. We consider $NPS\_LTR\_score^{k,l}_{i} = \{\emptyset\}$ when a user with that role does not evaluate alternative $A_i$ with this test. 

 In order to collect these values in the $ID^{k,l}$ matrix, we transform the non-empty scores, \textit{i.e.}, the LTR values obtained through the NPS test, to the $S_{SUS}$ scale. In order to do so, we perform two steps:
 
\begin{enumerate}

    \item Transform NPS values $NPS\_LTR\_score^{k,l}_{i}$ into $NPS\_SUS\_score^{k,l}_{i}$ values. Sauro \textit{et al}~\cite{sauro2016quantifying} establish a relationship between the values provided by users to the question posed in the NPS test (LTR values) and SUS values. The first proposed approach provides the equation $LTR = SUS/10$ for predicting the LTR value through a SUS value in a very simple way. Subsequently, through a study with more than 2000 users, the authors were able to obtain a better approximation of the regression equation that relates both values as $LTR = 1.33 + 0.08(SUS)$. We use the latter approximation since it is more accurate. Therefore, a correspondence is established between the scores obtained after answering the LTR question of the NPS test and a SUS test by:

        \begin{equation}
        \label{NPS_LTR}
        NPS\_SUS\_score^{k,l}_{i} = \dfrac{NPS\_LTR\_score^{k,l}_{i}-1.33}{0.08}
        \end{equation}
        
    If the value $NPS\_SUS\_score^{k,l}_{i}$ is negative, then it is reset to 0. If that value is greater than 100, we set it to 100.
    
    \item Transform the NPS values $NPS\_SUS\_score^{k,l}_{i}$ in \modif{2-tuples} \modif{associated to} $S_{SUS}$ and collect them. We establish a correspondence between the values obtained in the NPS test represented as scores of a SUS test to the linguistic scale $S_{SUS}$ by means of Definition \ref{def:SUS}. We transform the scores $NPS\_SUS\_score^{k,l}_{i}, i=1, \dots, n$ of user $U_k$ playing the role $R_l$ into linguistic 2-tuples associated with the $S_{SUS}$ scale and compile them into $ID^{k,l}$ matrices. More specifically, for each user $U_k$ playing the role $R_l$, we complete the second column ($j=2$) of its associated $ID^{k,l}$ matrix as follows:
    \begin{equation}\label{eq:TFNPS}
        ID^{k,l}_{i2} = TF_{SUS}(NPS\_SUS\_score^{k,l}_{i}) \; \; \forall i=1, \dots, n
    \end{equation}
    \end{enumerate}
    
 \item [$C3 \cong UT$.] While the user $U_k$ is playing the role $R_{l}$ and performing the set of tasks, we collect some useful information: $Success\_score^{k,l}_{i}(q_v)$ (score on the success of the task), $Efficiency\_score^{k,l}_{i}(q_v)$ (score on the efficiency in time in the task) and an assessment regarding the user experience performing the UT, the $Satisfaction\_score^{k,l}_{i} \in S^5$. The first two are a value score which is assigned to a given $q_v$ task and included as percentages in the final report, while satisfaction is considered as a linguistic variable for the test $C_3$. More specifically, for each user $U_k$ playing the role $R_{l}$, we complete the third column ($j=3$) of its associated matrix $ID^{k,l}$ as follows: 
    \begin{equation}
    \label{eq:pu}
        ID^{k,l}_{i,3} = (Satisfaction\_score^{k,l}_i,0) \; \;  \forall i=1,\dots,n        
    \end{equation}

   \item [$C4 \cong ACC$.] Since this test labels each alternative $A_i$ as A, AA or AAA, the result already belongs to $S^3$ and is stored as a 2-tuple. For each expert user $U_k$ playing the role $R_l$, we then complete the fourth column ($j=4$) of its associated $ID^{k,l}$ matrix as follows:
   \begin{equation}
        ID^{k,l}_{i4} = (Acc\_score^{k,l}_{i}, 0) \; \; \forall i=1, \dots, n
    \end{equation}

\end{description}

In summary, we highlight the linguistic scales of the elements of the $ID^{k,l}$ matrices, represented as 2-tuples:

\begin{itemize}
    \item $ID^{k,l}_{i1} \leadsto S_{SUS}$
    \item $ID^{k,l}_{i2} \leadsto S_{SUS}$
    \item $ID^{k,l}_{i3} \leadsto S^{5}$
    \item $ID^{k,l}_{i4} \leadsto S^{3}$
\end{itemize}

\subsection{Phase 4. Collective aggregation}
\label{sub:agregacion}

In this phase, the individual ratings of the users playing different roles collected in the $ID^{k,l}$ matrices are aggregated into $r+1$ collective matrices: one matrix for each role $R_l, (l=1,\dots,r)$ and a global matrix that aggregates the matrices of each role.

\bigskip
\label{unificacion}\textbf{Step 9. Unification of the information to $S^9$}. We must unify the values of the individual matrices in order to aggregate them. We transform each matrix $ID^{k,l} \neq \{\emptyset\}$ containing the original user evaluations $U_k$ playing the role $R_{l}$ into a matrix with the unified evaluations in $S^{9}$, the deepest level of the hierarchy $LH$. We call these the \textit{unified individual decisions }($UID$) matrices. Depending on the linguistic scale used in the evaluation of each criterion, one transformation or another must be applied. \modif{We present} how to proceed with each one:

\begin{itemize}
    \item $C_1$ and $C_2$. Rates are on the scale $S_{SUS}$. We use the transformation function $\mathscr{LH}$ (see Section V.A of \cite{herrera2008fuzzy}) to transform 2-tuples of an unbalanced linguistic scale, such as $S_{SUS}$, into a linguistic hierarchy, such as $LH = S^3 \cup S^5 \cup S^9$. In particular, we set $\mathscr{LH}$:$(S_{SUS} \times [-0.5, 0.5))$ $\rightarrow$ $(LH \times [-0.5, 0.5))$. 
    After this conversion, the linguistic assessments are expressed in different linguistic domains which means that they cannot be processed directly. We require the transformation function $\mathscr{TF}$ (see Section V.B of \cite{herrera2008fuzzy}) to convert the 2-tuples from different domains into a particular granularity label set of $LH$. We set $\mathscr{TF}_{3}^{t}$: $(LH \times [-0.5, 0.5))$ $\rightarrow$ $(S^{9} \times [-0.5, 0.5))$ as a special function that integrates a set of transformation functions $TF$~\cite{herrera2001} between levels of $LH$ to the highest level of $LH$. In our case, by the definition of SUS, the obtained transformed 2-tuples can belong to either $S^5$ or $S^9$, thereby resulting in:
    \begin{equation}
    \label{eq:unifc12}
        UID^{k,l}_{ij} = \mathscr{TF}_{3}^{t}(\mathscr{LH}(ID^{k,l}_{ij})) \; \; \forall  j=1,2; i=1, \dots, n
    \end{equation}

    with $t$ being level 2 or 3 of the hierarchy. If $ID^{k,l}_{ij} = \{\emptyset\}$, then $UID^{k,l}_{ij} = \{\emptyset\}$. \modif{We omit $\mathscr{LH}$ and $\mathscr{TF}$ for extension (fully available in \cite{herrera2008fuzzy}).}      

    \item $C_3$ and $C_4$. Rates are on the scale $S^5$ and $S^3$\modif{, respectively}. We use the transformation function $TF$~\cite{herrera2001}, which enables to transform 2-tuples between any level of a linguistic hierarchy into $LH$. More specifically, the third and fourth columns of the $UID^{k,l}$ matrices are obtained by transforming the 2-tuples of the $ID^{k,l}$ matrices to $S^9$ as follows:
    \begin{equation}
    \label{eq:unifc34}
    UID^{k,l}_{ij} = TF_{3}^{t}(ID^{k,l}_{ij}) = \Delta \left( \frac{\Delta^{-1}(ID^{k,l}_{it}) \times 8}{n(t)-1} \right)  \; \; \forall  j=3,4; i=1, \dots, n
    \end{equation}
    
    with $t$ being level 1 or 2 of the hierarchy. If $ID^{k,l}_{ij} = \{\emptyset\}$, then $UID^{k,l}_{ij} = \{\emptyset\}$.
    
\end{itemize}

\bigskip
\label{sub:agregacionxuxr}\textbf{Step 10. Aggregation for each role}. We define the \textit{unified collective decision matrix for role $R_l$} ($UCD^l$) containing the unified collective decisions in $S^9$ including the unified individual decisions of all users with role $R_l$. We obtain $UCD^{l}$ as the aggregation of non-empty $UID^{k,l}$, $(k=1,\dots,u)$ matrices by means of the 2TWA operator. Each element $UCD^{l}_{ij}$, ($i=1,\dots,n; j=1,\dots,m$) is defined as:


\begin{equation}   \label{eq:xuser} 
    \begin{split}
    UCD^{l}_{ij} =& 2TWA_{W^l}(UID^{1,l}_{ij}, \dots, UID^{u,l}_{ij}) \\
            =& \Delta \left( \dfrac{\sum_{k=1}^{u}{\Delta^{-1}(UID^{k,l}_{ij}) \times W_{k}^{l}}}{\sum_{k=1}^{u}{W_{k}^{l}}}  \right) = (s^{l}_{ij}, \alpha^{l}_{ij})
    \end{split}
\end{equation}

where the elements of the vector of weights $W^{l} = (W^{l}_{1}, \dots, W^{l}_{u})$ for role $R_{l}$ are defined by $W^{l}_{k} = \frac{W'^{l}_{k}}{\sum_{k=1}^{u}{W'^{l}_{k}}}, k=1, \dots, u$, such as:
\begin{equation}\label{eq:pesos}
\begin{split}
W'^{l}_{k} =
        &\begin{cases}
        WU_{k} \hspace{0.7cm} \; \; if \; \; UID^{k,l} \neq \{\emptyset\}\\
        0  \hspace{1.5cm} if \; \; UID^{k,l} = \{\emptyset\}
        &\end{cases}
\end{split}
\end{equation}

We continue by aggregating based on the criteria weights. This process generates a \textit{unified collective decision (\textbf{ucd}$^l)$} vector for each role $R_l$, which is used in the usability report. 

\begin{equation}\label{eq:xcriteria}
    \textbf{ucd}_{i}^{l} = \Delta \left ( \Delta^{-1}(s_{ij}^{l},\alpha_{ij}^{l}) \times WC_j \right ).
\end{equation}

\bigskip
\label{sub:agregacionxr}\textbf{Step 11. Global aggregation.} A \textit{unified collective decision $UCD^{global} n\times m$} matrix containing the unified collective decisions in $S^9$ is defined by aggregating the unified collective decisions of each role $R_l$. For this purpose, the 2TWA operator is applied as the aggregation of the non-empty $UCD^{l}_{ij}$, $(l=1,\dots,r)$ matrices. Each element of $UCD^{global}_{ij}$, ($i=1,\dots,n$; $j=1,\dots,m$) is defined by Equation \ref{eq:xrol}:

\begin{equation}\label{eq:xrol}
    \begin{split}
        UCD^{global}_{ij} =&  2TWA_{WR}(UCD^{1}_{ij}, \dots, UCD^{r}_{ij}) \\
        =& \Delta \left(\sum_{l=1}^{r}{\Delta^{-1}(UCD^{l}_{ij}) \times WR_{l}} \right)   = (s^{global}_{ij}, \alpha^{global}_{ij})
    \end{split}
\end{equation}

where $WR$ is the normalized vector of weights of the roles. Aggregation is then performed based on the criteria weights. This process results in the generation of a global unified collective decision (\textbf{ucd}$^{global}$) vector, which is then used in the usability report.

\begin{equation}\label{eq:global}
    \textbf{ucd}^{global}_{i} = \Delta \left ( \Delta^{-1}(s^{global}_{ij},\alpha^{global}_{ij}) \times WC_j \right ).
\end{equation}

\subsection{Phase 5. Data exploitation} 
\label{sub:ranking}
We apply the TOPSIS method (Section \ref{sec:topsismethod}) on the unified collective decisions matrices to generate several rankings of the alternatives in order to derive a ranking for specific roles and a general ranking. Thus, the model builds $r+1$ \textit{rankings}: one for each role $R_l$ based on matrices $UCD^{l}, (l=1,\dots,r)$ and a global ranking based on matrix $UCD^{global}$. The ranking of the alternatives is established according to the relative closeness coefficient to the ideal alternative. The higher the coefficient value, the better the alternative $A_i$ is. All alternatives $A_i (i=1,2, \dots, m)$ can then be ranked according to a descending order of the relative closeness values.

\bigskip
\textbf{Step 12. Generation of rankings for each role.} The TOPSIS procedure is applied on the matrices $UCD^{l}$, $(l=1,\dots,r)$ whose values are 2-tuples $(s^{l}_{ij}, \alpha^{l}_{ij})$. Therefore, we set $(s_{ij}, \alpha_{ij})$ = $(s^{l}_{ij}, \alpha^{l}_{ij})$ to obtain $r$ rankings, one for each role, which we denote by $Ranking^{l}, (l=1, \dots, r)$.

\bigskip
\textbf{Step 13. Global ranking generation.} The TOPSIS procedure is applied on the $UCD^{global}$ matrix whose values are 2-tuples $(s^{global}_{ij}, \alpha^{global}_{ij})$. Therefore, we set $(s_{ij}, \alpha_{ij})$ = $(s^{global}_{ij}, \alpha^{global}_{ij})$ to obtain the global ranking that we denote by $Ranking^{global}$.

\bigskip
\textbf{Step 14. Retranslation.} The elements of the vectors $\textbf{ucd}^{l}$, $(l=1,\dots,r)$, and $\textbf{ucd}^{global}$ are 2-tuples in $S^9$. This step conducts a retranslation of these values to $S_{SUS}$ in order to provide more comprehensible linguistic terms for the users. It is not mandatory to perform this step to apply the LDM4WUE methodology, but it is convenient to complete the usability report.

We build an \textit{adjective usability report} ($\textbf{aur}^{l}$) vector for each role $R_l$ and a global \textit{adjective usability report} ($\textbf{aur}^{global}$) vector containing the information from the $\textbf{ucd}^{l}$ and $\textbf{ucd}^{global}$ vectors, respectively, as linguistic terms of $S_{SUS}$. First, we define an identity function $id$ to transform 2-tuple linguistic terms from $S^9$ to the linguistic hierarchy $LH = S^3 \cup S^5 \cup S^9$ by $id(s, \alpha) = (s, \alpha)$. We set $id$:$(S^9 \times [-0.5, 0.5))$ $\rightarrow$ $(LH \times [-0.5, 0.5))$. Subsequently, we use the inverse transformation function $\mathscr{LH}^{-1}$ (see Section V.A of \cite{herrera2008fuzzy}) which transforms the linguistic 2-tuple expressed in a linguistic hierarchy, such as $LH$, into an unbalanced linguistic scale, such as $S_{SUS}$. We then set $\mathscr{LH}^{-1}$:$(LH \times [-0.5, 0.5))$ $\rightarrow$ $(S_{SUS} \times [-0.5, 0.5))$. We compute the vectors $\textbf{aur}^{l}$, $(l=1,\dots,r)$ and $\textbf{aur}^{global}$ ($i=1,\dots,n; j=1,\dots,m$) using Equations \ref{eq:retranslationRoles} and \ref{eq:retranslationGlobal}, respectively.
\begin{equation}
    \label{eq:retranslationRoles}
        \textbf{aur}^{l}_{i} = \mathscr{LH}^{-1}(id(\textbf{ucd}^{l}_{i})) \; \; \forall  l=1, \dots, r
\end{equation}   
\begin{equation}
    \label{eq:retranslationGlobal}
        \textbf{aur}^{global}_{i} = \mathscr{LH}^{-1}(id(\textbf{ucd}^{global}_{i})) 
\end{equation}

\section{Online A/B testing DSS-based tool for web usability assessment}\label{sec:soft}

The LDM4WUE methodology stands out for its flexibility in combining the various tests that a designer must face when doing accessibility and usability assessment. The moderator is the name used for the person in charge of validating the best design among the alternatives established in the A/B test. This person would benefit from using a tool to assist this process, especially if it were free and available for online use. 

We provide an online DSS for web usability evaluation called \modif{USE-AB-DSS} (available at \url{https://lionware.dev/use}). It enables to set up the project parameters, share forms to users (experts and end-users) and enables us to quickly obtain the usability report of the conducted A/B test. \modif{Section~\ref{sub:project} provides details about how to set up a project and Section~\ref{sub:report} presents how the final results are obtained.}


\subsection{Project configuration under \modif{USE-AB-DSS}}\label{sub:project}

\begin{figure}[H]
\centering
     \includegraphics[width=12cm]{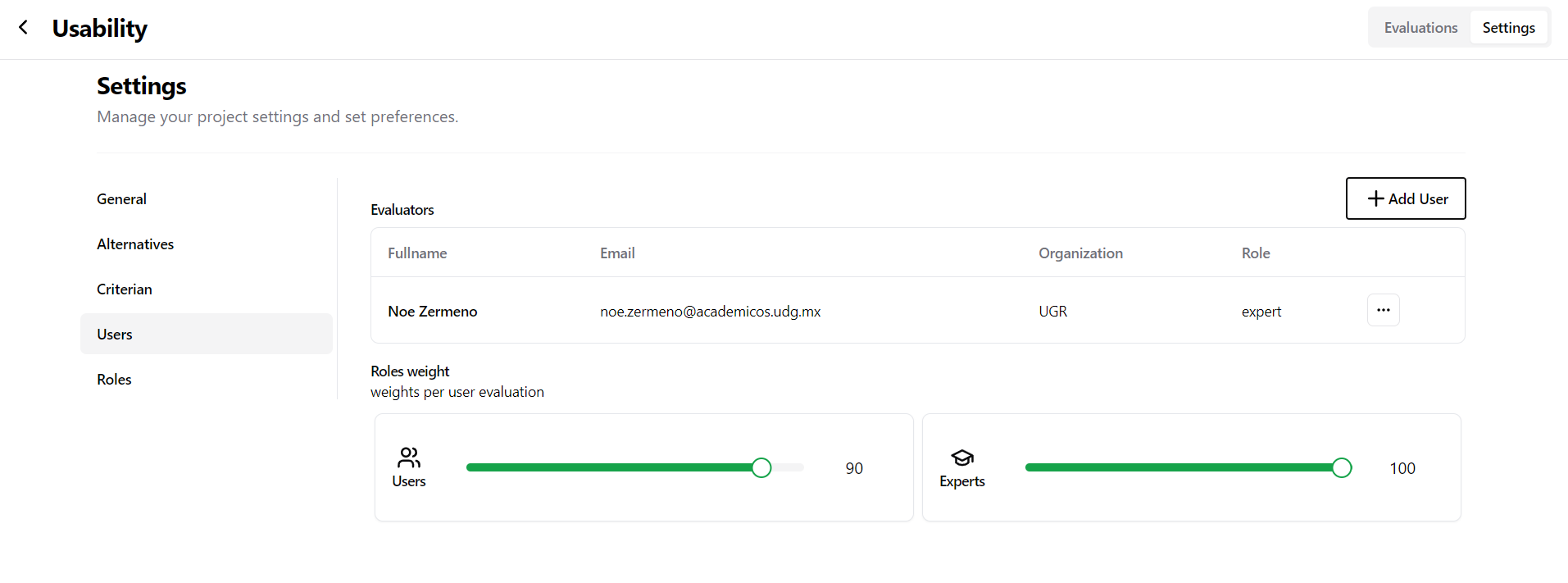}
    \caption{Configuration page to set the users. Here the relative importance is 100\% for experts and 90\% for the end-users.}
    \label{fig:user-set}
\end{figure}

\begin{figure}[H]
\centering
     \includegraphics[width=12cm]{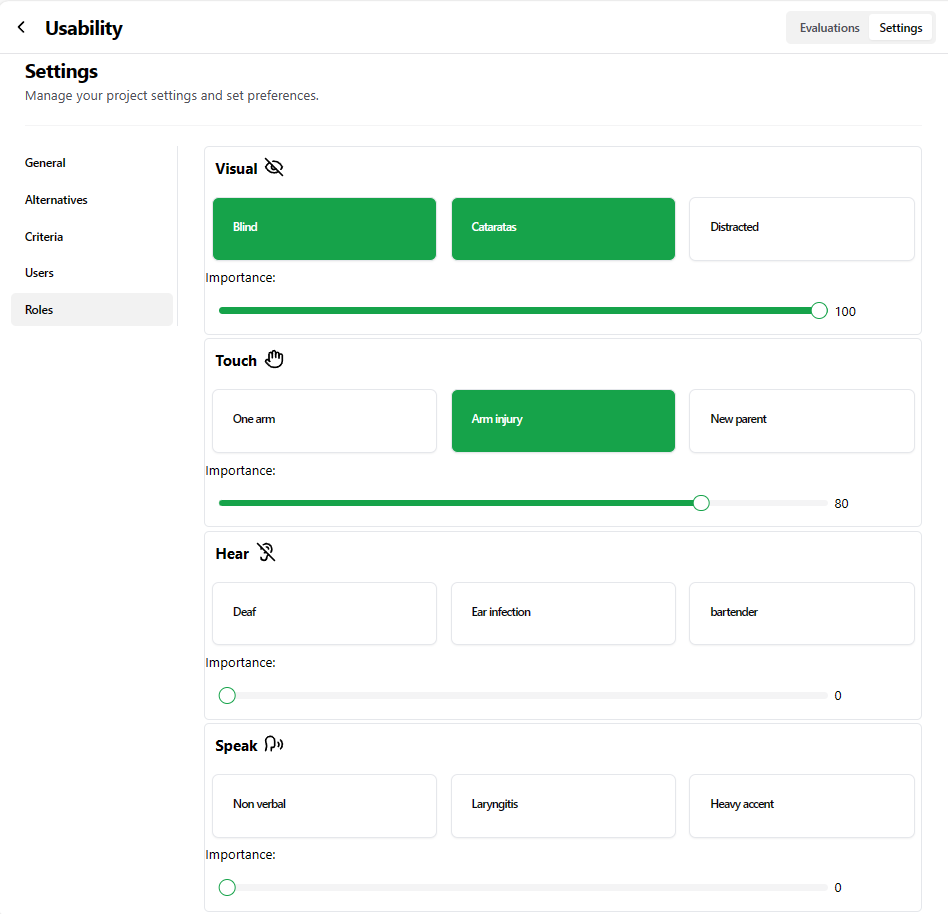}
    \caption{Configuration page to set the role-playing possibilities.}
    \label{fig:roles-set}
\end{figure}

Before conducting a usability test or any other inspection procedure, it is necessary to provide certain information about the websites to be evaluated. The following steps are, therefore, used to create a project:
\begin{enumerate}
    \item \textbf{\modif{Alternatives} settings}. After creating the project template, the moderator adds each alternative for evaluation using the `Add' icon. The moderator then enters a short name, the URL, and optionally, the website logo image for each alternative.

    \item \textbf{Criteria settings}. The moderator selects questionnaires from the test set in \textcolor{blue}{USE-AB-DSS} DSS to evaluate the alternatives. New tests can be added from the \textit{System Test} menu so that any UT can be configured with specific tasks targeted to the functionality of the websites to be evaluated. Linguistic labels indicating the importance of criteria in the comparison matrix are then chosen. A consistency index \modif{lower than }$ .10$ confirms the validity of criteria importance, enabling the evaluation project configuration to proceed.

    \item \textbf{Users settings}. The moderator \modif{set} a group of experts \modif{and a group of end-users} and specifies \modif{the weight of each group as shown in Figure~\ref{fig:user-set}.}

    \item \textbf{\modif{Roles} settings}. The moderator selects disabilities from predefined system categories. For instance, if they identify that people with visual and touch disabilities may access the sites to be evaluated, only those two role categories could be selected. Furthermore, it is possible to assign a different importance to each selected role simply by using a slider. Figure \ref{fig:roles-set} shows the selection of some roles and their weights.

\end{enumerate}

\subsection{Conducting the A/B testing with the \modif{USE-AB-DSS}}\label{sub:report}

\begin{figure}[H]
\centering
    \includegraphics[width=12cm]{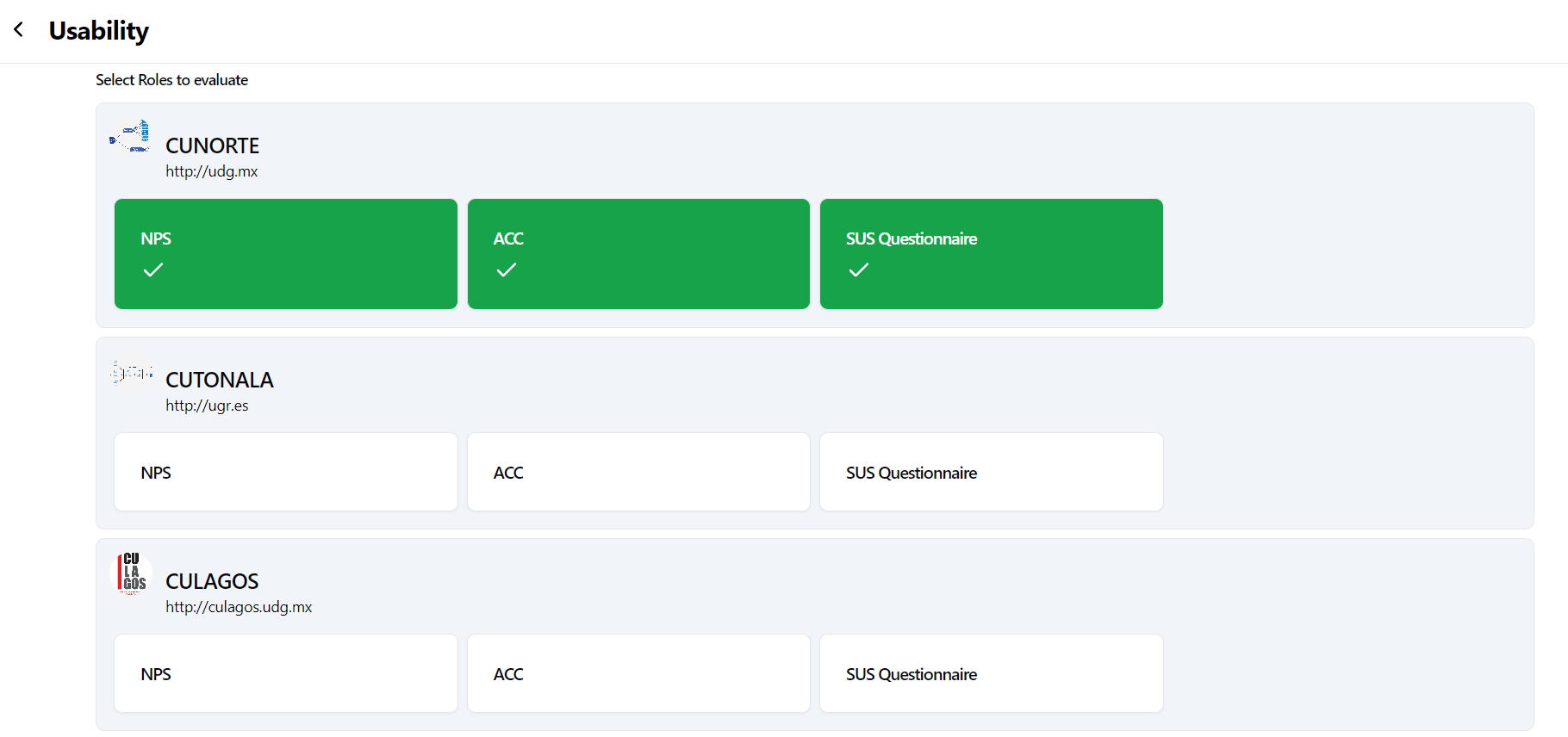}
    \caption{Dashboard view of experts and recipients to evaluate three alternatives through the NPS, SUS and ACC tests.}
    \label{fig:user-anwsers}
\end{figure}

\begin{figure}[H]
\centering
     \includegraphics[width=12cm]{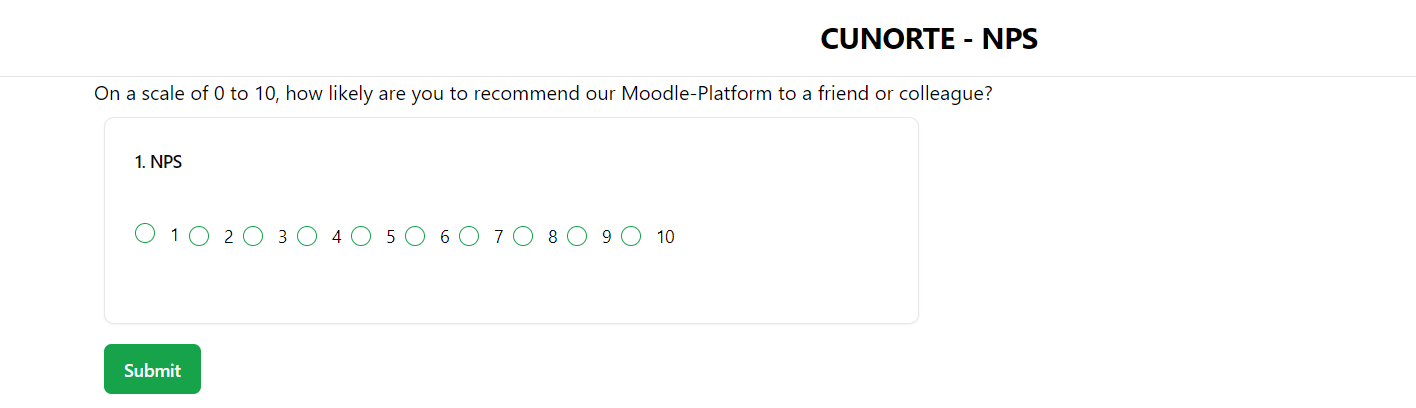}
    \caption{NPS - answer.}
    \label{fig:nps}
\end{figure}

Once the participants are enrolled on a usability evaluation project and have a role to play, they select the tests that evaluate each of the alternatives. A check will show which tests have been performed and which are still pending. Figure~\ref{fig:user-anwsers} shows the \modif{USE-AB-DSS} evaluation dashboard with three tests completed for the first alternative. Tests are completed by our tool. For instance, Figure~\ref{fig:nps} is the interface that completes the NPS test.

\begin{figure}[H]
\centering
     \includegraphics[width=14.5cm]{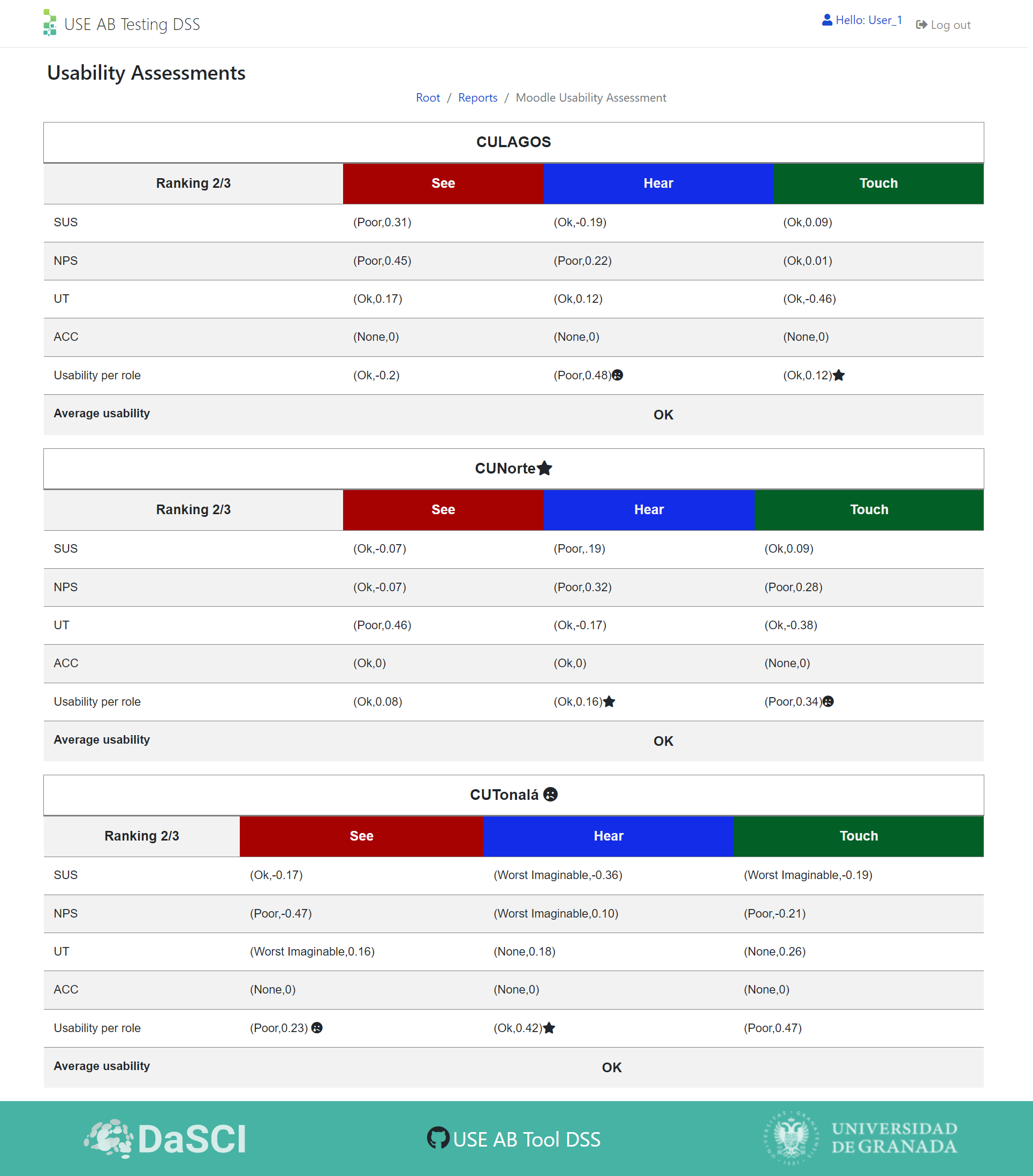}
    \caption{Usability evaluation report for three Moodle platforms obtained with \modif{USE-AB-DSS}.}
    \label{fig:report}
\end{figure}

Finally, \modif{the USE-AB-DSS} computes the steps from 8 to 14 and \modif{gives} the results of the A/B test in the reporting menu. Although our case of study is described in the following section, the reporting can be seen in Figure~\ref{fig:report}.

\section{Case of study: virtual learning environments usability evaluation}\label{sec:case}

In the decision-making field, real case studies and practical tools are essential for promoting and facilitating application in various contexts~\cite{cabrerizo15,Pam18,adepoju2019survey,ambarwati2021usability,Agrawal22,Eldrandaly23}. With the increase in the number of people working from home or taking online classes, new challenges regarding the use of technology are emerging. One of the major threats in terms of technologies is loss of interest in \textit{design for all} and adaptability to student needs, since blended or hybrid education scenarios are increasingly common in Higher Education Institutions (HEI). These new scenarios, particularly in the context of teaching and learning~\cite{buenano2019use}, need to be assessed. Any HEI wishing to provide an inclusive virtual environment should focus on three aspects:
\begin{enumerate}
    \item Usability. This focuses on the platform and should be as useful as possible, and this is particularly relevant when there is a wide group of users such as for a \modif{MOOC.}

    \item Educational methodologies. These focus on the contents and materials that teachers share with their students and should be designed on the basis of the Universal Design for Learning (UDL) paradigm~\cite{DUA11}.

    \item Inclusive aspects. These enhance the system with assistive technologies (AT) which are designed to help groups of people with special needs. 
\end{enumerate}


\modif{ We propose the testing of a popular learning management systems (LMS), the one known as Moodle.\footnote{About Moodle LMS \url{https://docs.moodle.org/402/en/About_Moodle}}}

\subsection{Case \modif{of study} - Phase 1: problem \modif{description}}

We present a case use for the LDM4WUE methodology given in Section~\ref{sec:model}. We aim to assess the usability of three Moodle learning platforms. These sites correspond to three separate Universities centers, with different installed Moodle versions with different features and themes. They all share, however, the same course content which is to be used in the usability test.

\begin{figure}[H]
    \centering
    \begin{subfigure}[c]{.49\textwidth}
        \centering
        \includegraphics[width=\textwidth]{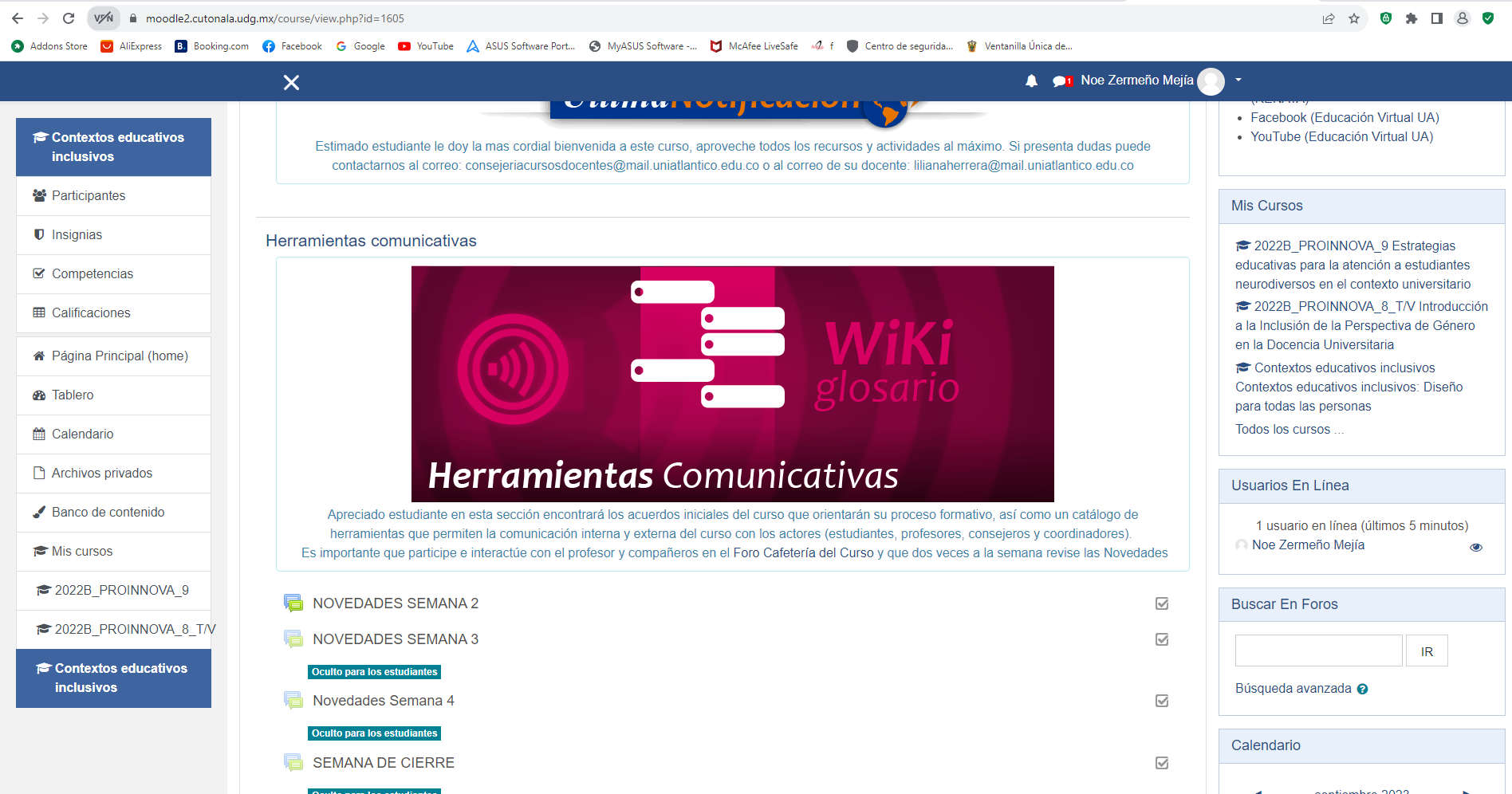}
        \caption{$A_1=$ DUA-MOOC at CUTonalá.}
        \label{fig:menu}
    \end{subfigure}
    \hfill
    \begin{subfigure}[c]{.49\textwidth}
        \centering
        \includegraphics[width=\textwidth]{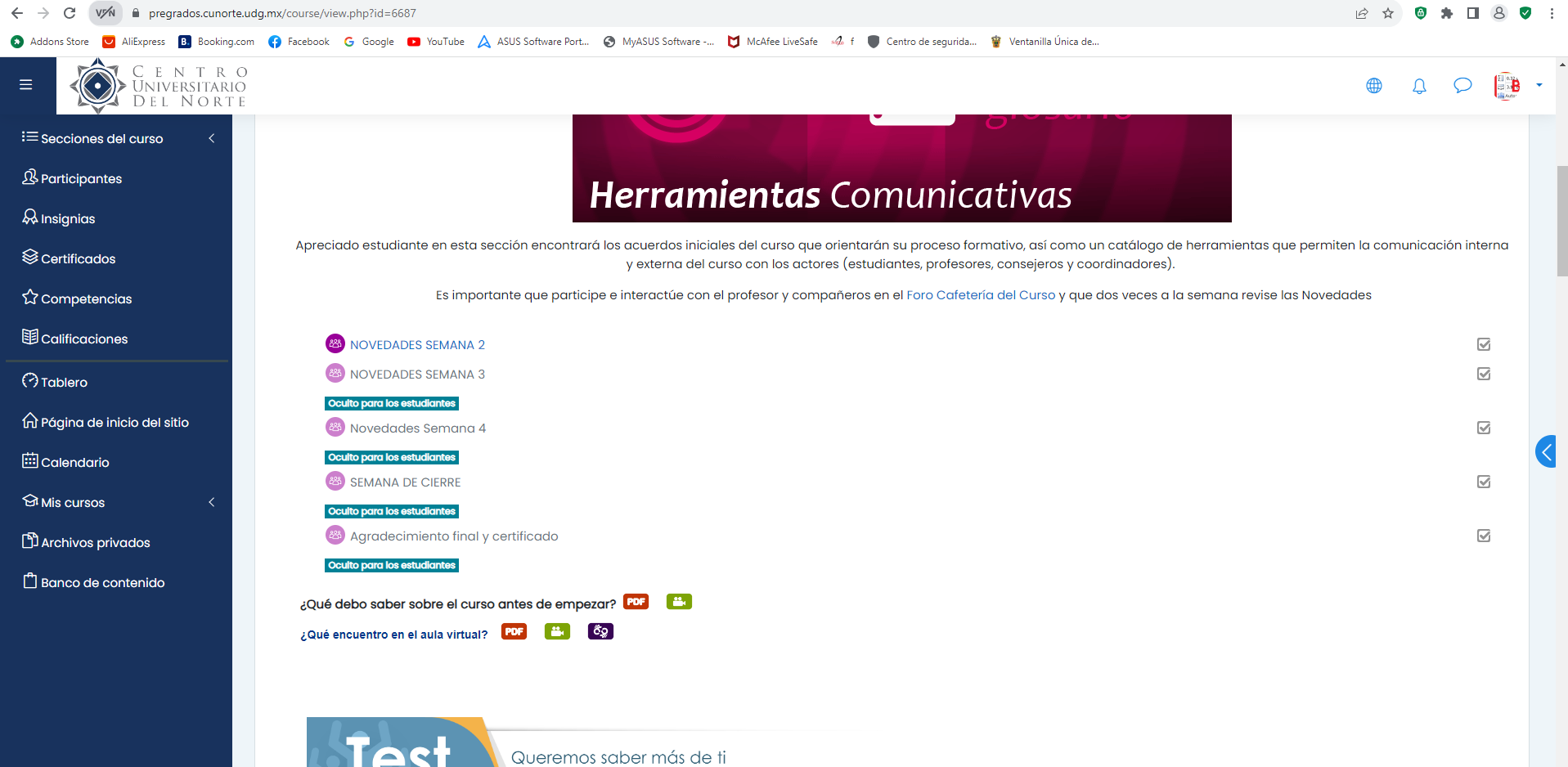}
        \caption{$A_2=$ DUA-MOOC at CUNorte.}
        \label{fig:filtro}
    \end{subfigure}
    \begin{subfigure}[c]{.49\textwidth}
        \centering
        \includegraphics[width=\textwidth]{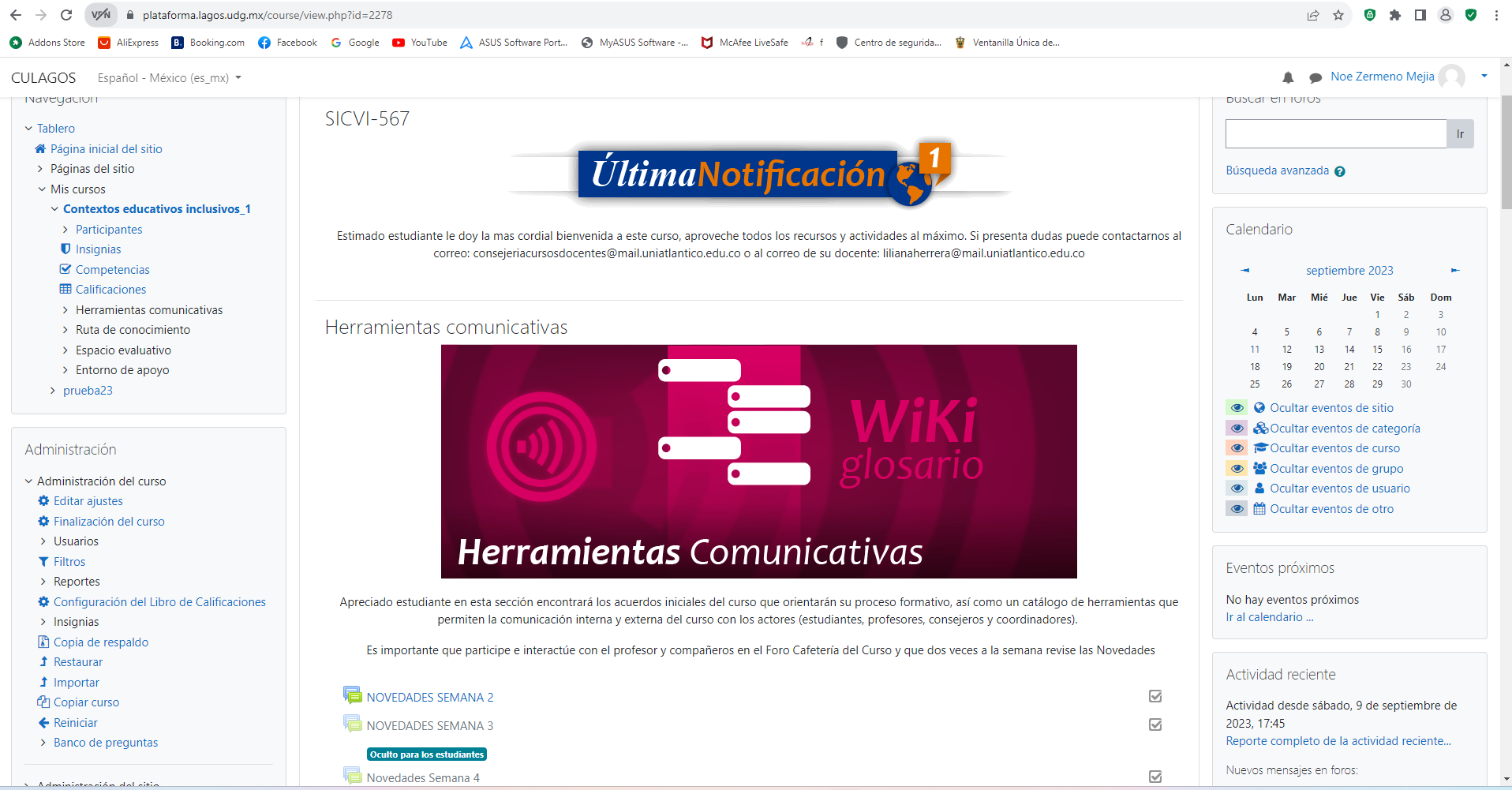}
        \caption{$A_3=$ DUA-MOOC at CULagos.}
        \label{fig:metrics}
    \end{subfigure}
    \caption{Using the same course on each website set the focus of the evaluation into the MOODLE theme.}
    \label{fig:moodle}
\end{figure}

\bigskip
\textbf{Step 1. Definition of the alternative set}. Three websites are established as the set of alternatives: the University of Guadalajara Tonalá Center Moodle platform\footnote{CUTonala \url{https://moodle2.cutonala.udg.mx/course/view.php?id=1605}} (CUTonala), the University of Guadalajara Northern University Center Virtual Campus\footnote{CUNorte \url{https://pregrados.cunorte.udg.mx/course/view.php?id=6687}} (CUNorte), and the University of Guadalajara Center of Los Lagos Moodle platform\footnote{CULagos \url{https://plataforma.lagos.udg.mx/course/view.php?id=2278}} (CULagos). Our set of alternatives are, therefore, $A = \{\mbox{CULagos}, \mbox{CUNorte}, \mbox{CUTonala}\}$. Figure~\ref{fig:moodle} shows the DUA-MOOC main entry at each alternative website.

\bigskip
\textbf{Step 2. Definition of the criteria set and derivation of the weight vector}. Let $C = \{\mbox{SUS}, \mbox{NPS}, \mbox{UT}, \mbox{ACC}\}$ be the set of criteria through which the usability of each alternative $A_i$ is evaluated. More specifically for $C_3$, the moderator must define the content of the usability test and the estimated maximum time per task. The publicly available \textit{Usability Test for Moodle under Universal Design for Learning}\footnote{OD-Moodle-Usability-Assessment at GitHub \url{https://github.com/ari-dasci/OD-Moodle-Usability-Assessment}} defines $UT=\{q_1, \dots, q_{28}\}$. The requested tasks to be performed are listed in Table~\ref{tbl:ut_task}. We subsequently determine the criteria weights.

\begin{table}[h]
\center
\scalebox{0.8}{
\begin{tabular}{ll}
\hline
\multicolumn{1}{c}{\textbf{Task category}} & \multicolumn{1}{c}{\textbf{Task list}} \\ \hline
\multirow{3}{*}{Log in to the platform} & 1. Log in to Moodle  \\ \cline{2-2} 
      & 2. Find a course                                                        \\ \cline{2-2} 
      & 3. Access the course                                                            \\ \hline
\multirow{3}{*}{Technical support access}
    & 4. Find technical support documentation (manual, FAQ)                                            \\ \cline{2-2}                             
    & 5. Complete the technical support contact form     \\ \cline{2-2}                              
    & 6. Switch site language                                               \\ \hline
\multirow{2}{*}{User account management} 
    & 7. Edit your profile                                                              \\ \cline{2-2}       
    & 8. Upload/Update profile photo                                               \\ \hline
\multirow{8}{*}{\begin{tabular}[c]{@{}l@{}}Access to information and \\ resources/content\end{tabular}} 
    & 9. Read news items in What's new \\ \cline{2-2} 
    & 10. Download a file                                                      \\ \cline{2-2} 
    & 11. Download a file from the resource Directory                             \\ \cline{2-2} 
    & 12. Track an external URL link to the platform                           \\ \cline{2-2}
    & 13. Display an embedded video                                            \\ \cline{2-2} 
    & 14. View a Page resource                                                      \\ \cline{2-2} 
    & 15. On the Page: read the text                                                  \\ \cline{2-2} 
    & 16. On the Page: visualize an image                                       \\ \hline
\multirow{2}{*}{Communication}           
    & 17. Send a message to a co-worker/teacher                                \\ \cline{2-2} 
     & 18. Participate in a Chat                          \\ \hline
\multirow{10}{*}{\begin{tabular}[c]{@{}l@{}}Accomplishment of course activities \end{tabular}}      
    & 19. Upload a file in the Task resource                                         \\ \cline{2-2} 
    & 20. Answer a Questionnaire resource                                                \\ \cline{2-2} 
    & 21. Add an entry to the Glossary resource                                           \\ \cline{2-2} 
    & 22. Select a sub-group                                                           \\ \cline{2-2} 
    & 23. Participate in the Forum resource                                                \\ \cline{2-2} 
    & 24. In the editor: format text  \\ \cline{2-2} 
    & 25. In the editor: insert a new link                                                        \\ \cline{2-2} 
    & 26. In the editor: insert an image                                                       \\ \cline{2-2} 
    & 27. In the editor: zoom in or out (change size or make full screen) \\ \cline{2-2} 
    & 28. Track your grades                                  \\ \hline
\end{tabular}
}
    \caption{Task list for the Moodle Usability Test under the UDL paradigm. Source:~\cite{liliana20}, pp 159.}
    \label{tbl:ut_task}

\end{table}

\bigskip
\textbf{Step 2.1. Obtain pairwise judgments regarding the importance of criteria and complete the criteria comparison matrix $CP$}. The moderator is responsible for determining the weight of each criterion $C_j$. \modif{This person} begin by establishing the importance of one criterion over another, using a linguistic assessment to express the relative importance of each criterion test. Table~\ref{fahp:t1} details the moderator's assessment of each pair of criteria. We transform the linguistic labels into TFNs and complete the information of the $CP$ matrix by using Equation~\ref{completarCP} (see Table~\ref{fahp:matcp'}).

\begin{table}[h]
\center
\scalebox{0.9}{
\begin{tabular}{rllllllllllll}
\hline
\multicolumn{1}{c}{\textbf{}} & \multicolumn{3}{c}{\textbf{$C_1$}} & \multicolumn{3}{c}{\textbf{$C_2$}}          & \multicolumn{3}{c}{\textbf{$C_3$}}    & \multicolumn{3}{c}{\textbf{$C_4$}} \\ \hline
\textbf{$C_1$}                & \multicolumn{3}{l}{Just import.} & \multicolumn{3}{l}{Very strongly import.} & \multicolumn{3}{l}{Equally import.} & \multicolumn{3}{l}{Weak import.} \\
\textbf{$C_2$}                & \multicolumn{3}{l}{}               & \multicolumn{3}{l}{Just import.}          & \multicolumn{3}{l}{Equally import.} & \multicolumn{3}{l}{Just import.} \\
\textbf{$C_3$}                & \multicolumn{3}{l}{}               & \multicolumn{3}{l}{}                        & \multicolumn{3}{l}{Just import.}    & \multicolumn{3}{l}{Weak import.} \\
\textbf{$C_4$}                & \multicolumn{3}{l}{}               & \multicolumn{3}{l}{}                        & \multicolumn{3}{l}{}                  & \multicolumn{3}{l}{Just important} \\ \hline
\end{tabular}}
\caption{Assessment of the $CP$ matrix by the moderator.}
\label{fahp:t1}
\end{table}

\begin{table}[h]
\center
\scalebox{0.9}{
\begin{tabular}{rcccccccccccc}
\hline
\multicolumn{1}{l}{} & \multicolumn{3}{c}{\textbf{$C_1$}}                                                                                          & \multicolumn{3}{c}{\textbf{$C_2$}}                                                                                          & \multicolumn{3}{c}{\textbf{$C_3$}}                                                                                          & \multicolumn{3}{c}{\textbf{$C_4$}}                                                                                          \\ \hline
\multicolumn{1}{l}{} & \multicolumn{1}{r}{\textit{\textbf{l}}} & \multicolumn{1}{r}{\textit{\textbf{m}}} & \multicolumn{1}{r}{\textit{\textbf{u}}} & \multicolumn{1}{r}{\textit{\textbf{l}}} & \multicolumn{1}{r}{\textit{\textbf{m}}} & \multicolumn{1}{r}{\textit{\textbf{u}}} & \multicolumn{1}{r}{\textit{\textbf{l}}} & \multicolumn{1}{r}{\textit{\textbf{m}}} & \multicolumn{1}{r}{\textit{\textbf{u}}} & \multicolumn{1}{r}{\textit{\textbf{l}}} & \multicolumn{1}{r}{\textit{\textbf{m}}} & \multicolumn{1}{r}{\textit{\textbf{u}}} \\
\textbf{$C_1$}       & 1                                       & 1                                       & 1                                       & 5                                       & 7                                       & 9                                       & 1                                       & 1                                       & 3                                       & 1                                       & 3                                       & 5                                       \\
\textbf{$C_2$}       & $^1/_9$                                     & $^1/_7$                                      & $^1/_5$                                     & 1                                       & 1                                       & 1                                       & 1                                       & 1                                       & 3                                       & 1                                       & 1                                       & 1                                       \\
\textbf{$C_3$}       & $^1/_3$                                      & 1                                       & 1                                       & $^1/_3$                                     & 1                                       & 1                                       & 1                                       & 1                                       & 1                                       & 1                                       & 3                                       & 5                                       \\
\textbf{$C_4$}       & $^1/_5$                                    & $^1/_3$                                     & 1                                       & 1                                       & 1                                       & 1                                       & $^1/_5$                                     & $^1/_3$                                     & 1                                       & 1                                       & 1                                       & 1                                       \\ \hline
\end{tabular}
}
\caption{The $CP$ matrix is set with the moderator's rates for each criterion.}
\label{fahp:matcp'}
\end{table}

\bigskip
\textbf{Step 2.2. Computing the fuzzy synthetic extension}. The fuzzy synthetic extension is calculated using Equations \ref{eq:extsintetica1}, \ref{eq:extsintetica2} and \ref{eq:extsintetica3}. See Table~\ref{fahp:exension} for the expression that is derived in this step.
\begin{table}[!hbt]
\center
\scalebox{0.9}{
\begin{tabular}{@{}llll@{}}
                  & \multicolumn{3}{c}{\textbf{Eq.~\ref{eq:extsintetica2}}}                                                               \\
\textbf{Criteria} & \multicolumn{1}{c}{\textbf{l}} & \multicolumn{1}{c}{\textbf{m}} & \multicolumn{1}{c}{\textbf{u}} \\
\hline
\textbf{$C_1$}    & 8.00                           & 12.00                          & 18.00                          \\
\textbf{$C_2$}    & 3.11                           & 3.14                           & 5.20                           \\
\textbf{$C_3$}    & 2.67                           & 6.00                           & 8.00                           \\
\textbf{$C_4$}    & 2.40                           & 2.67                           & 4.00                           \\
\multirow{4}{*}{} & \multicolumn{3}{c}{\textbf{Eq.~\ref{eq:extsintetica3}}}    \\
\hline
                  & 0.028                          & 0.042                          & 0.062                          \\
                  & \multicolumn{3}{c}{\textbf{Eq.~\ref{eq:extsintetica1}}}\\
                  & \multicolumn{1}{c}{\textbf{l}} & \multicolumn{1}{c}{\textbf{m}} & \multicolumn{1}{c}{\textbf{u}} \\
\hline                  
\textbf{$C_1$}    & 0.23                           & 0.50                           & 1.11                           \\
\textbf{$C_2$}    & 0.09                           & 0.13                           & 0.32                           \\
\textbf{$C_3$}    & 0.08                           & 0.25                           & 0.49                           \\
\textbf{$C_4$}    & 0.07                           & 0.11                           & 0.25                          
\end{tabular}
}
\caption{Fuzzy synthetic extension calculation.}
\label{fahp:exension}
\end{table}

\bigskip
\textbf{Step 2.3. Possibility Degree}. The degree of possibility is calculated using Equation~\ref{eq:degreeofpossibility}, resulting in the values in Table \ref{fahp:possibility}.

\begin{table}[h]
\center
\scalebox{0.9}{
\begin{tabular}{rcrcrcrc}
\hline
\multicolumn{1}{c}{\textbf{$C_1 >   C_j$}} & \textbf{PD} & \multicolumn{1}{c}{\textbf{$C_2 > C_j$}} & \textbf{PD} & \multicolumn{1}{c}{\textbf{$C_3 > C_j$}} & \textbf{PD} & \multicolumn{1}{c}{\textbf{$C_4 > C_j$}} & \textbf{PD} \\ \hline
\textbf{SC1\textgreater{}SC2}              & 1.00        & \textbf{SC2\textgreater{}SC1}            & 0.20        & \textbf{SC3\textgreater{}SC1}            & 0.51        & \textbf{SC4\textgreater{}SC1}            & 0.05        \\
\textbf{SC1\textgreater{}SC3}              & 1.00        & \textbf{SC2\textgreater{}SC3}            & 0.67        & \textbf{SC3\textgreater{}SC2}            & 1           & \textbf{SC4\textgreater{}SC2}            & 0.89        \\
\textbf{SC1\textgreater{}SC4}              & 1.00        & \textbf{SC2\textgreater{}SC4}            & 1.00        & \textbf{SC3\textgreater{}SC4}            & 1           & \textbf{SC4\textgreater{}SC3}            & 0.55        \\ \hline
\end{tabular}
}
\caption{Calculus of the possibility degree.}
\label{fahp:possibility}
\end{table}

\label{WC}\textbf{Step 2.4. Obtaining the vectors of weights for the set $C$.} The vector of weights $WC' = \{1, 0.222, 0.515, 0.048\}$ is obtained using Equation \ref{eq:weight}. Finally, Equation \ref{eq:pesonorm} is used to derive the normalized weight vector of the criteria 
$$
WC = \{0.567, 0.114, 0.292, 0.027\}.
$$

We know the moderator's assessment regarding the relative importance of criteria is a correct evaluation using the consistency index ($CI$). In this scenario, $CI  = -.08 \leq .10$ is valid~\cite{Rodrigues14}. Otherwise, the moderator is prompted to change the $CP$ matrix.

\bigskip
\textbf{Step 3. Definition of the user set}. Let $E = \{E_1, \dots, E_4\}$ be the set of experts and $D = \{D_1, \dots, D_{11}\}$ the set of end-users, and let $U= \{U_1,\dots, U_{15}\}$ be the union of experts and end-users, such that $U = E \cup D$. Each user $U_k$ evaluated the usability of each alternative $A_i$ through each criterion $C_j$.

Furthermore, it is assumed that the expert set $E$ has more knowledge in the UX discipline of \modif{HCI} so that their opinion can carry more weight in the general usability assessment than that given by $D$. In any case, it is up to the moderator to decide to change this policy for another of interest, for example, when participants are users with real disabilities and they want to increase the weight of this collective and so we allow the moderator to express this particular importance for each group. For this case, we have $WE=100\%$ and $WD=90\%$, and considering the membership of each user to one of these two groups, the following vector is stored. 
$$WU = \{1, 1, 1, 1, .9, .9, .9, .9, .9, .9, .9, .9, .9, .9,
.9, .9, .9, .9, .9\}$$ 

By considering the role selected by the users, we can then calculate the corresponding normalized vector of user weights $W^l$ using Equation \ref{eq:pesos}.

\bigskip
\textbf{Step 4. Definition of the set of roles}. The moderator selects the following roles $R = \{$\textit{Blind}, \textit{Ear infection}, \textit{Arm injury}$\}$ as the set of possible choices, or equivalently, as the maximum number of times that a given user can answer the A/B test by playing roles without ever repeating. The moderator then assigns importance to the roles by setting the role weight vector $WR' = \{100, 75, 75\},$ which is normalized to \label{WR}$WR = \{0.40, 0.30, 0.30\}$.

It is crucial to emphasize that users commence with the usability test ($C_3$) and are subsequently free to choose any other test. This consideration is associated with the time spent on the website during the usability test, as this criterion can be the most exploratory test for alternatives. It is therefore extremely useful since it experiments with system usability.

Knowing which role each user plays allows us to calculate the normalization of the user weights for each role. Our report confirms that users $U_4$, $U_6$ and $U_{12}$ selected $R_1$ as they felt they wanted to play that role. Users $U_2$, $U_5$, $U_{10}$, $U_{13}$ and $U_{15}$ played $R_2$. Finally, users $U_1$, $U_3$, $U_7$, $U_8$, $U_9$, $U_{11}$ and $U_{14}$ selected role $R_3$ to play. When a user selects a role, this is maintained for every test and for each alternative. People are free to select any other role and restart the A/B testing, answering from another point of view/need each test for every alternative. 


By using Equation~\ref{eq:pesos} on $WU$ for each role $R_l$ $(l=1,2,3)$, we derive the normalized user weights.
$$W^{1}=\{0, 0, 0, 0.357, 0, 0.321, 0, 0, 0, 0, 0, 0.321, 0, 0, 0\}$$
$$W^{2}=\{0, 0.217, 0, 0, 0.196, 0, 0, 0, 0, 0.196, 0, 0, 0.196, 0, 0.196\}$$
$$W^{3}=\{0.154, 0, 0.154, 0, 0, 0, 0.138, 0.138, 0.138, 0, 0.138, 0, 0, 0.138, 0\}$$

\subsection{Case \modif{of study} - Phases 2 and 3: empathy and data elicitation}

We have conducted this A/B testing with real people on the \textit{Graphics, Interfaces and Usability} (GIU) course. GIU is a fourth year course of the Computer Science Degree at the University of Guadalajara, Mexico. For this case use, the full set of answers (450 responses for SUS, 45 responses for NPS, 1260 responses for the UT and 12 responses for ACC) can be downloaded from the project repository at GitHub.\footnote{\textit{USE-AB-DSS}. \url{https://github.com/ari-dasci/S-USE-AB-Tool}} In order to understand how the LDM4UWE methodology uses this information in a meaningful way, we move onto Phase 2 and subsequent phases.

\bigskip
\textbf{Steps 5 and 6. Briefing and Role Playing}. In practice, the people who participate in a UT need a brief introduction to let them know in advance what they will be asked to do. The methodology proposes three standardized tests and the task list from Table~\ref{tbl:ut_task}. The test leader has also determined that the group of expert users are the teachers on the GUI course. The students act as end-users. The tests are proposed to the students as practical exercises to carry out accessibility and usability tests. The full A/B testing (assessments for each alternative and every test) took less than two hours, including the time spent giving instructions, choosing roles, and two short breaks.

\bigskip
\textbf{Step 7. Gathering user evaluations}. Tests are executed sequentially, either by using various data collection instruments or assisted by software. We need to run each $C_j$ and gather user responses, and adapt these to our linguistic decision-making model. Answers to the SUS Questionnaire ($C_1$) are presented in Table~\ref{tbl:C1}. The NPS ratings (test $C_2$) are shown in Table~\ref{tbl:C2}. Full input data for $C_3$ test is not given here due to space restrictions. Instead, we provide Table~\ref{tbl:UT_single} with the information gathered from user $U_4^1$. This shows the efficiency and success metrics per task of this particular user when running the UT for each alternative. A quick look shows that $U_4^1$ performed best with $A_3$. Finally, and only for the group of experts $D$, the automatic accessibility assessment reports are linguistically interpreted in terms of the number of errors and warnings. This information is provided in Table~\ref{tbl:C4}.

\begin{table}[h]
\center
\scalebox{0.8}{
\begin{tabular}{l|ccccccccccccccc}
\multicolumn{1}{c}{} & \multicolumn{1}{l}{\textbf{$U_1^{3}$}} & \multicolumn{1}{l}{\textbf{$U_2^2$}} & \multicolumn{1}{l}{\textbf{$U_3^{3}$}} & \multicolumn{1}{l}{\textbf{$U_4^1$}} & \multicolumn{1}{l}{\textbf{$U_5^{2}$}} & \multicolumn{1}{l}{\textbf{$U_6^1$}} & \multicolumn{1}{l}{\textbf{$U_7^{3}$}} & \multicolumn{1}{l}{\textbf{$U_8^3$}} & \multicolumn{1}{l}{\textbf{$U_9^{3}$}} & \multicolumn{1}{l}{\textbf{$U_{10}^2$}} & \multicolumn{1}{l}{\textbf{$U_{11}^{3}$}} & \multicolumn{1}{l}{\textbf{$U_{12}^1$}} & \multicolumn{1}{l}{\textbf{$U_{13}^{2}$}} & \multicolumn{1}{l}{\textbf{$U_{14}^3$}} & \multicolumn{1}{l}{\textbf{$U_{15}^{2}$}} \\
\hline 
\textbf{$A_1$}       & 42.5                                 & 30                                   & 52.5                                 & 60                                   & 50                                   & 30                                   & 62.5                                 & 60                                   & 65                                   & 32.5                                  & 30                                    & 37.5                                  & 47.5                                  & 40                                    & 42.5                                  \\
\textbf{$A_2$}       & 50                                   & 55                                   & 62.5                                 & 42.5                                 & 80                                   & 57.5                                 & 55                                   & 47.5                                 & 62.5                                 & 50                                    & 60                                    & 45                                    & 57.5                                  & 62.5                                  & 47.5                                  \\
\textbf{$A_3$}      & 60                                   & 27.5                                 & 75                                   & 30                                   & 40                                   & 27.5                                 & 50                                   & 55                                   & 72.5                                 & 50                                    & 57.5                                  & 37.5                                  & 57.5                                  & 60                                    & 70                       
\end{tabular}
}
\caption{Input view of $C_1 \cong SUS$ responses for each $A_i$.} \label{tbl:C1}
\end{table}

\begin{table}[h]
\center
\scalebox{0.9}{
\begin{tabular}{@{}l|ccccccccccccccc@{}}
\toprule
 &
  \multicolumn{1}{l}{\textbf{$U_{1}^3$}} &
  \multicolumn{1}{l}{\textbf{$U_{2}^2$}} &
  \multicolumn{1}{l}{\textbf{$U_{3}^3$}} &
  \multicolumn{1}{l}{\textbf{$U_{4}^1$}} &
  \multicolumn{1}{l}{\textbf{$U_{5}^2$}} &
  \multicolumn{1}{l}{\textbf{$U_{6}^1$}} &
  \multicolumn{1}{l}{\textbf{$U_{7}^3$}} &
  \multicolumn{1}{l}{\textbf{$U_{8}^3$}} &
  \multicolumn{1}{l}{\textbf{$U_{9}^3$}} &
  \multicolumn{1}{l}{\textbf{$U_{10}^2$}} &
  \multicolumn{1}{l}{\textbf{$U_{11}^3$}} &
  \multicolumn{1}{l}{\textbf{$U_{12}^1$}} &
  \multicolumn{1}{l}{\textbf{$U_{13}^2$}} &
  \multicolumn{1}{l}{\textbf{$U_{14}^3$}} &
  \multicolumn{1}{l}{\textbf{$U_{15}^2$}} \\ \midrule
\textbf{$A_1$}  & 4 & 4 & 6 & 3 & 6 & 6 & 5 & 5 & 7 & 6 & 7 & 6 & 5 & 1 & 6 \\
\textbf{$A_2$}  & 7 & 7 & 1 & 2 & 8 & 7 & 5 & 8 & 6 & 8 & 8 & 7 & 5 & 2 & 8 \\
\textbf{$A_3$} & 3 & 3 & 5 & 1 & 5 & 7 & 1 & 8 & 5 & 8 & 7 & 7 & 1 & 7 & 6 \\ \bottomrule
\end{tabular}%
}
\caption{Input view of $C_2 \cong NPS$ with $NPS\_LTR_{score}$ data for each $A_i$.}
\label{tbl:C2}
\end{table}

\begin{table}[h]
\centering
\scalebox{0.5}{
\begin{tabular}{@{}cc||cccc|cccc|cccc@{}}
\toprule
\textbf{UT} &
  \multicolumn{1}{l}{\textbf{}} &
  \multicolumn{4}{c}{$A_1 = \{\mbox{CULagos}\}$} &
  \multicolumn{4}{c}{$A_2 = \{\mbox{CUNorte}\}$} &
  \multicolumn{4}{c}{$A_3 = \{\mbox{CUTonalá}\}$} \\ \midrule
\textbf{$q_v$} &
  \multicolumn{1}{l}{\textbf{$MaxTime$}} &
  \textbf{$Time^{4,1}_1$} &
  \textbf{$Effic.^{4,1}_1$} &
  \textbf{$Success^{4,1}_1$} &
  \textbf{$Satisf.^{4,1}_1$} &
  \textbf{$Time^{4,1}_2$} &
  \textbf{$Effic.^{4,1}_2$} &
  \textbf{$Success^{4,1}_2$} &
  \textbf{$Satisf.^{4,1}_2$} &
  \textbf{$Time^{4,1}_3$} &
  \textbf{$Effic.^{4,1}_3$} &
  \textbf{$Success^{4,1}_3$} &
  \textbf{$Satisf.^{4,1}_3$} \\
  \midrule
\textbf{$q_1$}  & 5   & 5   & 1 & 1 & $s^5_3$ & 4   & 1 & 1 & $s^5_4$ & 6   & 0 & 1 & $s^5_1$ \\
\textbf{$q_2$}  & 20  & 16  & 1 & 1 & $s^5_4$ & 24  & 0 & 1 & $s^5_4$ & 29  & 0 & 1 & $s^5_0$ \\
\textbf{$q_3$}  & 10  & 11  & 0 & 1 & $s^5_4$ & 11  & 0 & 1 & $s^5_4$ & 12  & 0 & 1 & $s^5_1$ \\
\textbf{$q_4$}  & 30  & 38  & 0 & 0 & $s^5_0$ & 31  & 0 & 0 & $s^5_0$ & 21  & 1 & 1 & $s^5_2$ \\
\textbf{$q_5$}  & 30  & 32  & 0 & 0 & $s^5_0$ & 32  & 0 & 0 & $s^5_0$ & 31  & 0 & 0 & $s^5_0$ \\
\textbf{$q_6$}  & 30  & 25  & 1 & 1 & $s^5_3$ & 21  & 1 & 1 & $s^5_4$ & 37  & 0 & 0 & $s^5_0$ \\
\textbf{$q_7$}  & 120 & 112 & 1 & 1 & $s^5_3$ & 92  & 1 & 1 & $s^5_4$ & 155 & 0 & 1 & $s^5_1$ \\
\textbf{$q_8$}  & 30  & 36  & 0 & 1 & $s^5_3$ & 41  & 0 & 1 & $s^5_3$ & 42  & 0 & 0 & $s^5_0$ \\
\textbf{$q_9$}  & 120 & 131 & 0 & 1 & $s^5_3$ & 169 & 0 & 1 & $s^5_3$ & 135 & 0 & 0 & $s^5_0$ \\
\textbf{\textcolor{blue}{$q_{10}$}} & 30  & 34  & 0 & 0 & $s^5_0$ & 33  & 0 & 1 & $s^5_2$ & 31  & 0 & 0 & $s^5_0$ \\
\textbf{\textcolor{blue}{$q_{11}$}} & 45  & 58  & 0 & 0 & $s^5_0$ & 41  & 1 & 1 & $s^5_3$ & 61  & 0 & 0 & $s^5_0$ \\
\textbf{\textcolor{blue}{$q_{12}$}} & 30  & 30  & 0 & 1 & $s^5_2$ & 21  & 1 & 1 & $s^5_4$ & 41  & 0 & 1 & $s^5_2$ \\
\textbf{\textcolor{blue}{$q_{13}$}} & 120 & 113 & 1 & 1 & $s^5_2$ & 149 & 0 & 0 & $s^5_1$ & 124 & 0 & 0 & $s^5_0$ \\
\textbf{\textcolor{blue}{$q_{14}$}} & 45  & 36  & 1 & 1 & $s^5_4$ & 55  & 0 & 1 & $s^5_3$ & 55  & 0 & 0 & $s^5_0$ \\
\textbf{\textcolor{blue}{$q_{15}$}} & 120 & 134 & 0 & 1 & $s^5_1$ & 86  & 1 & 1 & $s^5_2$ & 135 & 0 & 0 & $s^5_0$ \\
\textbf{\textcolor{blue}{$q_{16}$}} & 20  & 26  & 0 & 1 & $s^5_1$ & 20  & 1 & 1 & $s^5_3$ & 25  & 0 & 0 & $s^5_0$ \\
\textbf{\textcolor{blue}{$q_{17}$}} & 45  & 60  & 0 & 1 & $s^5_3$ & 60  & 0 & 1 & $s^5_4$ & 50  & 0 & 0 & $s^5_0$ \\
\textbf{\textcolor{blue}{$q_{18}$}} & 90  & 84  & 1 & 1 & $s^5_1$ & 101 & 0 & 1 & $s^5_2$ & 123 & 0 & 0 & $s^5_0$ \\
\textbf{\textcolor{blue}{$q_{19}$}} & 90  & 85  & 1 & 1 & $s^5_2$ & 74  & 1 & 1 & $s^5_3$ & 122 & 0 & 0 & $s^5_0$ \\
\textbf{\textcolor{blue}{$q_{20}$}} & 600 & 769 & 0 & 0 & $s^5_2$ & 499 & 1 & 1 & $s^5_4$ & 621 & 0 & 0 & $s^5_0$ \\
\textbf{\textcolor{blue}{$q_{21}$}} & 180 & 182 & 0 & 1 & $s^5_3$ & 259 & 0 & 1 & $s^5_3$ & 249 & 0 & 1 & $s^5_1$ \\
\textbf{\textcolor{blue}{$q_{22}$}} & 60  & 77  & 0 & 0 & $s^5_1$ & 83  & 0 & 1 & $s^5_3$ & 85  & 0 & 0 & $s^5_0$ \\
\textbf{\textcolor{blue}{$q_{23}$}} & 60  & 54  & 1 & 1 & $s^5_2$ & 69  & 0 & 0 & $s^5_0$ & 69  & 0 & 1 & $s^5_2$ \\
\textbf{\textcolor{blue}{$q_{24}$}} & 120 & 110 & 1 & 1 & $s^5_3$ & 168 & 0 & 1 & $s^5_3$ & 138 & 0 & 1 & $s^5_0$ \\
\textbf{\textcolor{blue}{$q_{25}$}} & 30  & 40  & 0 & 0 & $s^5_0$ & 26  & 1 & 1 & $s^5_3$ & 32  & 0 & 0 & $s^5_0$ \\
\textbf{\textcolor{blue}{$q_{26}$}} & 30  & 23  & 1 & 1 & $s^5_3$ & 27  & 1 & 1 & $s^5_3$ & 33  & 0 & 1 & $s^5_0$ \\
\textbf{\textcolor{blue}{$q_{27}$}} & 45  & 57  & 0 & 0 & $s^5_0$ & 49  & 0 & 0 & $s^5_1$ & 51  & 0 & 1 & $s^5_1$ \\
\textbf{\textcolor{blue}{$q_{28}$}} & 60  & 52  & 1 & 1 & $s^5_2$ & 69  & 0 & 1 & $s^5_3$ & 62  & 0 & 1 & $s^5_0$ \\
\midrule
\textbf{Average} &
   &
   &
  \textbf{42.86\%} &
  \textbf{71.43\%} &
  \multicolumn{1}{c}{\textbf{$(s^5_2,-0.04)$}} &
  \textbf{} &
  \textbf{39.29\%} &
  \textbf{82.14\%} &
  \multicolumn{1}{c}{\textbf{$(s^5_3,-0.29)$}} &
  \textbf{} &
  \textbf{3.57\%} &
  \textbf{42.86\%} &
  \multicolumn{1}{c}{\textbf{$(s^5_0,0.39)$}} \\ \bottomrule
\end{tabular}%
}
\caption{A given user $U_4$ from $E$ set is running $C_3$ while playing role $R_1$.}\label{tbl:UT_single}
\end{table}

\begin{table}[h]
\center
\scalebox{0.9}{
\begin{tabular}{@{}l|cccc|ccccccccccc@{}}
\toprule
\multicolumn{6}{c}{\textbf{UX experts}} &
\multicolumn{8}{c}{\textbf{End-Users}}\\ \midrule
  $U_{1}^3$ &
  $U_{2}^2$ &
  $U_{3}^3$ &
  $U_{4}^1$ &
  $U_{5}^2$ &
  $U_{6}^1$ &
  $U_{7}^3$ &
  $U_{8}^3$ &
  $U_{9}^3$ &
  $U_{10}^2$ &
  $U_{11}^3$ &
  $U_{12}^1$ &
  $U_{13}^2$ &
  $U_{14}^3$ &
  $U_{15}^2$ \\ \midrule
$A_1$ & $A$ & $A$  & $A$ & $A$  & - & - & - & - & - & - & - & - & - & - & - \\
$A_2$ & $A$ & $AA$ & $A$ & $AA$ & - & - & - & - & - & - & - & - & - & - & - \\
$A_3$ & $A$ & $A$  & $A$ & $A$  & - & - & - & - & - & - & - & - & - & - & - \\
\bottomrule
\end{tabular}%
}
\caption{Input view of $C_4 \cong ACC$ with accessibility labels.}
\label{tbl:C4}
\end{table}

\textbf{Step 8. Construction of the individual decision matrices}. This step aims to construct the individual decision matrices with the answers of \modif{the} users. In order to perform linguistic collective aggregation, the data must be homogenized. For clarity purposes, we adhere to the procedure of user $U_4$ evaluating alternative $A_1$ while assuming the role $R_1$:
\begin{itemize}

    \item \textbf{Responses to $C_1 \cong SUS$}. The answers to the 10 items of the SUS questionnaire are $\{2, 3, 4, 2, 3, 2, 2, 2, 3, 1\}$. Therefore, using Equation \ref{eq:sus}, $SUS\_score^{4,1}_{1} = 60$. This numerical value is transformed using Equation \ref{eq:TFSus} to obtain $ID^{4,1}_{1,1}=TF_{SUS}(60)=(s^{sus}_2,0.4) \in S_{SUS}$. 

    \item \textbf{Responses to $C_2 \cong NPS$}. The direct response to the LTR question (how likely are you to recommend the LMS to an acquaintance or friend), is $4$. This numerical value is transformed into a linguistic 2-tuple using Equation \ref{NPS_LTR}, and thus  $NPS\_SUS_{score}= 33.375$. By applying Equation \ref{eq:TFNPS}, $ID^{4,1}_{1,2} = TF_{SUS}(33.375) = (s^{sus}_3,-0.335) \in S_{SUS}$.

    \item \textbf{Responses to $C_3 \cong UT$}. We apply a UT designed  to fully use an LMS environment (in~\cite{liliana20}, page 150). This comprises 28 tasks which are listed in Table~\ref{tbl:ut_task}. User $U_4^1$ gave the following results: 
    \begin{enumerate}
        \item Efficiency rate: out of 28 activities, 12 were completed with an efficiency rate of $Efficiency\_score^{4,1}_1 = 42.86$ (by Equation~\ref{eq:efficiency}).

        \item Success rate: out of the 28 activities, 20 were completed correctly obtaining $Success\_score^{4,1}_1 = 71.43$ (by Equation~\ref{eq:success}).
        \item Satisfaction level: satisfaction varied according to the task, but on average (by applying Equations~\ref{eq:satisfaction} and \ref{eq:pu}) we can derive for this user the $Satisfaction\_score^{4,1}_1 = (s^5_2,-0.036)$. Therefore, $ID^{4,1}_{1,3}$=$(s^{5}_2, -0.036) \in S^5$.
    \end{enumerate}

    \item \textbf{Responses to $C_4 \cong$ Accessibility}. Acting as an expert, User $U_4$ uses the WAVE tool to consult the LMS Moodle report and to obtain information about the evaluation of accessibility. This interpretation is summarized in the valuation given by label $A$. Therefore, $ID^{4,1}_{1,4}$ = $(A,0)=(s^3_0,0) \in S^3$.
    
\end{itemize}

Correspondingly, the elements of the $U_4^1$ individual decision matrix to evaluate the alternative $A_1$ are:
$$
ID^{4,1}_1 = \{(s^{sus}_2,0.4), (s^{sus}_3,-0.335), (s^{5}_2,-0.036),(s^3_0,0)\}
$$

Table~\ref{tbl:original} summarizes the matrices obtained in this step in relation to $U_k, (k=1, \dots,15)$. It should be noted that each $ID^{k,l}$ matrix is displayed in a single row of values, and that users have been grouped by roles.

\begin{table}[]
\center
\scalebox{0.5}{
\begin{tabular}{@{}l|llll|llll|llll@{}}
\toprule
  \multicolumn{1}{c}{}&
  \multicolumn{4}{c}{\textbf{$A_1 = \{CULagos\}$}} &
  \multicolumn{4}{c}{\textbf{$A_2 = \{CUNorte\}$}} &
  \multicolumn{4}{c}{\textbf{$A_3 = \{CUTonala\}$}} \\ \midrule
\textbf{$U_r^l$} &
  \textbf{$C_1 \cong SUS$} &
  \textbf{$C_2 \cong NPS$} &
  \textbf{$C_3 \cong UT$} &
  \textbf{$C_4 \cong ACC$} &
  \textbf{$C_1 \cong SUS$} &
  \textbf{$C_2 \cong NPS$} &
  \textbf{$C_3 \cong UT$} &
  \textbf{$C_4 \cong ACC$} &
  \textbf{$C_1 \cong SUS$} &
  \textbf{$C_2 \cong NPS$} &
  \textbf{$C_3 \cong UT$} &
  \textbf{$C_4 \cong ACC$} \\ \midrule
\textbf{$U_{4}^1$} &
  $(s^{sus}_2,0.4)$ &
  $(s^{sus}_3,-0.33)$ &
  $(s^5_2,-0.04)$ &
  $(s^3_0,0)$ &
  $(s^{sus}_3,0.4)$ &
  $(s^{sus}_3,-0.16)$ &
  $(s^5_3,-0.29)$ &
  $(s^3_1,0)$ &
  $(s^{sus}_2,0.4)$ &
  $(s^{sus}_1,-0.16)$ &
  $(s^5_0,0.39)$ &
  $(s^3_0,0)$ \\
\textbf{$U_{6}^1$} &
  $(s^{sus}_2,0.4)$ &
  $(s^{sus}_3,-0.33)$ &
  $(s^5_2,0.07)$ &
   &
  $(s^{sus}_2,0.3)$ &
  $(s^{sus}_3,-0.16)$ &
  $(s^5_2,0.21)$ &
   &
  $(s^{sus}_2,0.2)$ &
  $(s^{sus}_1,-0.16)$ &
  $(s^5_2,0.36)$ &
   \\
\textbf{$U_{12}^1$} &
  $(s^{sus}_3,0)$ &
  $(s^{sus}_2,0.33)$ &
  $(s^5_3,-0.5)$ &
   &
  $(s^{sus}_4,-0.4)$ &
  $(S^{sus}_0,0)$ &
  $(S^5_2,0.43)$ &
   &
  $(s^{sus}_3,0)$ &
  $(s^{sus}_4,-0.33)$ &
  $(s^5_3,-0.11)$ & 
   \\ \midrule
\textbf{$U_{2}^{2}$} &
  $(s^{sus}_2,0.4)$ &
  $(s^{sus}_1,-0.16)$ &
  $(s^5_2,-0.29)$ &
  $(s^3_0,0)$ &
  $(s^{sus}_2,0.2)$ &
  $(s^{sus}_0,0.33)$ &
  $(s^5_2,0.46)$ &
  $(s^3_1,0)$ &
  $(s^{sus}_2,0.2)$ &
  $(s^{sus}_0,0)$ &
  $(s^5_2,-0.14)$ &
  $(s^3_0,0)$ \\
\textbf{$U_{5}^2$} &
  $(s^{sus}_2,0)$ &
  $(s^{sus}_2,0.33)$ &
  $(s^5_3,0.14)$ &
   &
  $(s^{sus}_6,0.4)$ &
  $(s^{sus}_7,-0.33)$ &
  $(s^5_2,0.39)$ &
   &
  $(s^{sus}_3,0.2)$ &
  $(s^{sus}_4,-0.33)$ &
  $(s^5_3,-0.46)$ &
   \\
\textbf{$U_{10}^2$} &
  $(s^{sus}_3,-0.4)$ &
  $(s^{sus}_2,0.33)$ &
  $(s^5_3,-0.11)$ &
   &
  $(s^{sus}_2,0)$ &
  $(s^{sus}_3,-0.16)$ &
  $(s^5_2,0.32)$ &
   &
  $(s^{sus}_2,0)$ &
  $(s^{sus}_3,-0.16)$ &
  $(s^5_2,0.43)$ &
   \\
\textbf{$U_{13}^2$} &
  $(s^{sus}_4,-0.2)$ &
  $(s^{sus}_4,-0.33)$ &
  $(s^5_3,-0.25)$ &
   &
  $(s^{sus}_2,0.3)$ &
  $(s^{sus}_4,-0.33)$ &
  $(s^5_2,0.11)$ &
   &
  $(s^{sus}_2,0.3)$ &
  $(s^{sus}_0,0)$ &
  $(s^5_2,0.46)$ &
   \\
\textbf{$U_{15}^2$} &
  $(s^{sus}_3,0.4)$ &
  $(s^{sus}_4,-0.33)$ &
  $(s^5_3,-0.11)$ &
   &
  $(s^{sus}_4,-0.2)$ &
  $(s^{sus}_7,-0.33)$ &
  $(s^5_2,-0.11)$ &
   &
  $(s^{sus}_3,-0.2)$ &
  $(s^{sus}_7,-0.33)$ &
  $(s^5_2,-0.29)$ &
   \\ \midrule
\textbf{$U_{1}^{3}$} &
  $(s^{sus}_3,0.4)$ &
  $(s^{sus}_3,-0.16)$ &
  $(s^5_2,0.25)$ &
  $(s^3_0,0)$ &
  $(s^{sus}_2,0)$ &
  $(s^{sus}_2,0.33)$ &
  $(s^5_2,0.18)$ &
  $(s^3_0,0)$ &
  $(s^{sus}_2,0.4)$ &
  $(s^{sus}_4,-0.33)$ &
  $(s^5_3,-0.29)$ &
  $(s^3_0,0)$ \\
\textbf{$U_{3}^{3}$} &
  $(s^{sus}_2,0.1)$ &
  $(s^{sus}_2,0.33)$ &
  $(s^5_2,0.25)$ &
  $(s^3_0,0)$ &
  $(s^{sus}_3,-0.5)$ &
  $(s^{sus}_7,-0.33)$ &
  $(s^5_3,0.18)$ &
  $(s^3_0,0)$ &
  $(s^{sus}_3,0)$ &
  $(s^{sus}_7,-0.33)$ &
  $(s^5_3,-0.25)$ &
  $(s^3_0,0)$ \\
\textbf{$U_{7}^{3}$} &
  $(s^{sus}_3,-0.5)$ &
  $(s^{sus}_3,-0.16)$ &
  $(s^5_3,-0.46)$ &
   &
  $(s^{sus}_2,0.2)$ &
  $(s^{sus}_7,-0.33)$ &
  $(s^5_3,-0.43)$ &
   &
  $(s^{sus}_2,0)$ &
  $(s^{sus}_3,-0.16)$ &
  $(s^5_2,0.32)$ &
   \\
\textbf{$U_{8}^{3}$} &
  $(s^{sus}_2,0.4)$ &
  $(s^{sus}_2,0.33)$ &
  $(s^5_3,0.21)$ &
   &
  $(s^{sus}_4,-0.2)$ &
  $(s^{sus}_3,-0.16)$ &
  $(s^5_3,0.04)$ &
   &
  $(s^{sus}_2,0.2)$ &
  $(s^{sus}_3,-0.16)$ &
  $(s^5_3,-0.11)$ &
   \\
\textbf{$U_{9}^{3}$} &
  $(s^{sus}_3,-0.4)$ &
  $(s^{sus}_4,-0.33)$ &
  $(s^5_3,-0.11)$ &
   &
  $(s^{sus}_3,-0.5)$ &
  $(s^{sus}_4,-0.33)$ &
  $(s^5_2,0.32)$ &
   &
  $(s^{sus}_3,-0.1)$ &
  $(s^{sus}_0,0)$ &
  $(s^5_2,0.43)$ &
   \\
\textbf{$U_{11}^{3}$} &
  $(s^{sus}_2,0.4)$ &
  $(s^{sus}_0,0)$ &
  $(s^5_1,0.46)$ &
   &
  $(s^{sus}_2,0.4)$ &
  $(s^{sus}_0,0.33)$ &
  $(s^5_2,0.14)$ &
   &
  $(s^{sus}_2,0.3)$ &
  $(s^{sus}_3,-0.16)$ &
  $(s^5_3,0.14)$ &
   \\
\textbf{$U_{14}^{3}$} &
  $(s^{sus}_3,0.2)$ &
  $(s^{sus}_2,0.33)$ &
  $(s^5_3,0.21)$ &
   &
  $(s^{sus}_3,-0.5)$ &
  $(s^{sus}_7,-0.33)$ &
  $(s^5_3,-0.11)$ &
   &
  $(s^{sus}_2,0.4)$ &
  $(s^{sus}_2,0.33)$ &
  $(s^5_3,0.25)$ &
   \\ \bottomrule
\end{tabular}%
}
\caption{Elements of $ID^{k,l}$ matrices, represented as 2-tuples.}\label{tbl:original}
\end{table}

\subsection{Case \modif{of study -} Phase \modif{4}: collective aggregation}

In order to aggregate the information previously obtained from the set of users $U$, and considering the heterogeneity of the information, it was necessary to perform a unification process before aggregating the information. The necessary steps are detailed below.
\bigskip

\bigskip
\textbf{Step 9. Unification of the information to $S^9$}.\\

The values in Table~\ref{tbl:original} are unified to $S^9$ with the application of Equations~\ref{eq:unifc12} and~\ref{eq:unifc34}. The result of the aforementioned procedures for each alternative are shown in Table~\ref{tbl:s9}.

\begin{table}[h]
\center
\scalebox{0.5}{
\begin{tabular}{@{}l|llll|llll|llll@{}}
\toprule
 &
  \multicolumn{4}{c}{$A_1 = \{CULagos\}$} &
  \multicolumn{4}{c}{$A_2 = \{CUNorte\}$} &
  \multicolumn{4}{c}{$A_3 = \{CUTonala\}$} \\ \midrule
$U_r^l$ &
  $C_1 \cong SUS$ &
  $C_2 \cong NPS$ &
  $C_3 \cong UT$ &
  $C_4 \cong ACC$ &
  $C_1 \cong SUS$ &
  $C_2 \cong NPS$ &
  $C_3 \cong UT$ &
  $C_4 \cong ACC$ &
  $C_1 \cong SUS$ &
  $C_2 \cong NPS$ &
  $C_3 \cong UT$ &
  $C_4 \cong ACC$ \\ \midrule
$U_{4}^1$ &
  $(s^9_5,-0.2)$ &
  $(s^9_3,-0.33)$ &
  $(s^9_4,-0.07)$ &
  $(s^9_0,0)$ &
  $(s^9_3,0.4)$ &
  $(s^9_6,-0.32)$ &
  $(s^9_5,0.43)$ &
  $(s^9_4,0)$ &
  $(s^9_2,0.4)$ &
  $(s^9_2,-0.32)$ &
  $(s^9_1,-0.21)$ &
  $(s^9_0,0)$ \\
$U_{6}^1$ &
  $(s^9_2,0.4)$ &
  $(s^9_3,-0.33)$ &
  $(s^9_4,0.14)$ &
   &
  $(s^9_5,-0.4)$ &
  $(s^9_6,-0.32)$ &
  $(s^9_4,0.43)$ &
   &
  $(s^9_2,0.2)$ &
  $(s^9_2,-0.32)$ &
  $(s^9_5,-0.29)$ &
   \\
$U_{12}^1$ &
  $(s^9_3,0)$ &
  $(s^9_5,-0.34)$ &
  $(s^9_5,0)$ &
   &
  $(s^9_4,-0.4)$ &
  $(s^9_0,0)$ &
  $(s^9_5,-0.14)$ &
   &
  $(s^9_3,0)$ &
  $(s^9_4,-0.33)$ &
  $(s^9_6,-0.21)$ &
   \\ \midrule
$U_{2}^{2}$ &
  $(s^9_2,0.4)$ &
  $(s^9_2,-0.32)$ &
  $(s^9_3,0.43)$ &
  $(s^9_0,0)$ &
  $(s^9_4,0.4)$ &
  $(s^9_1,-0.34)$ &
  $(s^9_5,-0.07)$ &
  $(s^9_4,0)$ &
  $(s^9_2,0.2)$ &
  $(s^9_0,0)$ &
  $(s^9_4,-0.29)$ &
  $(s^9_0,0)$ \\
$U_{5}^2$ &
  $(s^9_4,0)$ &
  $(s^9_5,-0.34)$ &
  $(s^9_6,0.29)$ &
   &
  $(s^9_6,0.4)$ &
  $(s^9_7,-0.33)$ &
  $(s^9_5,-0.21)$ &
   &
  $(s^9_3,0.2)$ &
  $(s^9_4,-0.33)$ &
  $(s^9_5,0.07)$ &
   \\
$U_{10}^2$ &
  $(s^9_3,-0.4)$ &
  $(s^9_5,-0.34)$ &
  $(s^9_6,-0.21)$ &
   &
  $(s^9_4,0)$ &
  $(s^9_6,-0.32)$ &
  $(s^9_5,-0.36)$ &
   &
  $(s^9_4,0)$ &
  $(s^9_6,-0.32)$ &
  $(s^9_5,-0.14)$ &
   \\
$U_{13}^2$ &
  $(s^9_4,-0.2)$ &
  $(s^9_4,-0.33)$ &
  $(s^9_0,0)$ &
   &
  $(s^9_5,-0.4)$ &
  $(s^9_4,-0.33)$ &
  $(s^9_0,0)$ &
   &
  $(s^9_5,-0.4)$ &
  $(s^9_0,0)$ &
  $(s^9_5,-0.07)$ &
   \\
$U_{15}^2$ &
  $(s^9_3,0.4)$ &
  $(s^9_4,-0.33)$ &
  $(s^9_6,-0.21)$ &
   &
  $(s^9_4,-0.2)$ &
  $(s^9_7,-0.33)$ &
  $(s^9_4,-0.21)$ &
   &
  $(s^9_6,-0.4)$ &
  $(s^9_7,-0.33)$ &
  $(s^9_3,0.43)$ &
   \\ \midrule
$U_{1}^{3}$ &
  $(s^9_3,0.4)$ &
  $(s^9_6,-0.32)$ &
  $(s^9_5,-0.5)$ &
  $(s^9_0,0)$ &
  $(s^9_4,0)$ &
  $(s^9_5,-0.34)$ &
  $(s^9_4,0.36)$ &
  $(s^9_0,0)$ &
  $(s^9_5,-0.2)$ &
  $(s^9_4,-0.33)$ &
  $(s^9_5,0.43)$ &
  $(s^9_0,0)$ \\
$U_{3}^{3}$ &
  $(s^9_4,0.2)$ &
  $(s^9_5,-0.34)$ &
  $(s^9_5,-0.5)$ &
  $(s^9_0,0)$ &
  $(s^9_5,0)$ &
  $(s^9_7,-0.33)$ &
  $(s^9_6,0.36)$ &
  $(s^9_0,0)$ &
  $(s^9_6,0)$ &
  $(s^9_7,-0.33)$ &
  $(s^9_6,-0.5)$ &
  $(s^9_0,0)$ \\
$U_{7}^{3}$ &
  $(s^9_5,0)$ &
  $(s^9_6,-0.32)$ &
  $(s^9_5,0.07)$ &
   &
  $(s^9_4,0.4)$ &
  $(s^9_7,-0.33)$ &
  $(s^9_5,0.14)$ &
   &
  $(s^9_4,0)$ &
  $(s^9_6,-0.32)$ &
  $(s^9_5,-0.36)$ &
   \\
$U_{8}^{3}$ &
  $(s^9_5,-0.2)$ &
  $(s^9_5,-0.34)$ &
  $(s^9_6,0.43)$ &
   &
  $(s^9_4,-0.2)$ &
  $(s^9_6,-0.32)$ &
  $(s^9_6,0.07)$ &
   &
  $(s^9_4,0.4)$ &
  $(s^9_6,-0.32)$ &
  $(s^9_6,-0.21)$ &
   \\
$U_{9}^{3}$ &
  $(s^9_5,0.2)$ &
  $(s^9_4,-0.33)$ &
  $(s^9_6,-0.21)$ &
   &
  $(s^9_5,0)$ &
  $(s^9_4,-0.33)$ &
  $(s^9_5,-0.36)$ &
   &
  $(s^9_6,-0.2)$ &
  $(s^9_0,0)$ &
  $(s^9_5,-0.14)$ &
   \\
$U_{11}^{3}$ &
  $(s^9_2,0.4)$ &
  $(s^9_0,0)$ &
  $(s^9_3,-0.07)$ &
   &
  $(s^9_5,-0.2)$ &
  $(s^9_1,-0.34)$ &
  $(s^9_4,0.29)$ &
   &
  $(s^9_5,-0.4)$ &
  $(s^9_6,-0.32)$ &
  $(s^9_6,0.29)$ &
   \\
$U_{14}^{3}$ &
  $(s^9_3,0.2)$ &
  $(s^9_5,-0.34)$ &
  $(s^9_6,0.43)$ &
   &
  $(s^9_5,0)$ &
  $(s^9_7,-0.33)$ &
  $(s^9_6,-0.21)$ &
   &
  $(s^9_5,-0.2)$ &
  $(s^9_5,-0.34)$ &
  $(s^9_7,-0.5)$ &
   \\ \bottomrule
\end{tabular}%
} \caption{Unified Individual Decisions (UID) matrices expressed with $S^9$.} \label{tbl:s9}
\end{table}

\bigskip
\textbf{Step 10. Aggregation for each role}. Another useful piece of information is the unified collective decision vector, which is computed for each role. First, judgments in $S^9$ are clustered into a $UCD^l$ matrix for each role $R_l$. Let us follow the case of $l=1$ and users $U_k, k = \{4, 6, 12\}$ assessments represented in the $UCD^1$ matrix. We then apply Equation \ref{eq:xuser} by using the vector of user weights $W^1$ in order to derive $UCD^{l}_{ij}$ elements. Subsequently, by means of Equation \ref{eq:xcriteria} and the vector of criteria weights (see Step 2.4. at Section~\ref{WC}), we compute the unified collective decision \textbf{ucd}$^l$ vector per role. Results of both procedures are shown in Table \ref{tbl:ucd1}. This must then be  repeated to cover $UCD^l$ ($l = 2, \dots, r$).

\begin{table}[h]
\center
\scalebox{0.85}{
\begin{tabular}{@{}llllll@{}}
\toprule 
\multicolumn{1}{c}{\multirow{2}{*}{$A_i$}} &
  \multicolumn{1}{c}{$C_1$} &
  \multicolumn{1}{c}{$C_2$} &
  \multicolumn{1}{c}{$C_3$} &
  \multicolumn{1}{c}{$C_4$} &
  \multicolumn{1}{c}{\multirow{2}{*}{\textbf{ucd}$^{1}_{i}$}}\\
  \cmidrule{2-5}
\multicolumn{1}{c}{} &
  \multicolumn{1}{c}{$WC^1 = 0.567$} &
  \multicolumn{1}{c}{$WC^2 = 0.114$} &
  \multicolumn{1}{c}{$WC^3 = 0.292$} &
  \multicolumn{1}{c}{$WC^4 = 0.027$} &
  \multicolumn{1}{c}{} \\ \midrule
$A_1$  & $(s^9_3,0.45)$  & $(s^9_3,0.31)$  & $(s^9_4,0.34)$  & $(s^9_0,0)$ & $(s^9_4,-0.4)$  \\
$A_2$  & $(s^9_4,-0.15)$ & $(s^9_4,-0.15)$ & $(s^9_5,-0.08)$ & $(s^9_4,0)$ & $(s^9_4,0.17)$  \\
$A_3$ & $(s^9_3,-0.47)$ & $(s^9_2,0.32)$  & $(s^9_4,-0.34)$ & $(s^9_0,0)$ & $(s^9_3,-0.23)$ \\ \bottomrule
\end{tabular}%
} 
\caption{The $R^1$ role play enables us to derive the 2-tuple vector \textbf{ucd}$^1_i$ which contains the usability assessment for each alternative.} \label{tbl:ucd1}
\end{table}

\bigskip
\textbf{Step 11. Global aggregation}. In order to integrate all the information, we examine the complete matrix of $UCD^l_{ij}$ elements. Subsequently, we aggregate this information by taking into account the weights assigned to the roles, as outlined in Section~\ref{WR}, denoted as $WR$. We apply Equation~\ref{eq:xrol} to compute each global unified collective decision $UCD^{global}_{ij}$ element, and this is used to report a linguistic score for each usability test and for each alternative. 

Furthermore, a linguistic score can be derived to each alternative using Equation~\ref{eq:global}. We use \textbf{ucd}$^{global}_{i}$ to denote the unified collective decision vector and to represent website \textit{usability}. Table \ref{tbl:global} shows 2-tuples on $S^9$ that are the collective representation of the usability assessments given by all the users role-playing on the alternative websites through a series of tests. 

\begin{table}[h]
\center
\scalebox{0.9}{
\begin{tabular}{@{}llllll@{}}
\toprule
 & \textbf{ucd}$^{global}_{i1}$&
  \textbf{ucd}$^{global}_{i2}$ &
  \textbf{ucd}$^{global}_{i3}$ &
  \textbf{ucd}$^{global}_{i4}$ &
  \textbf{ucd}$^{global}_{i}$\\
\midrule
$A_1$  & $(s^9_4,-0.45)$ & $(s^9_4,-0.34)$ & $(s^9_5,-0.47)$ & $(s^9_0,0)$    & $(s^9_4,-0.24)$  \\
$A_2$  & $(s^9_4,0.30)$   & $(s^9_4,0.41)$  & $(s^9_5,-0.36)$ & $(s^9_3,-0.2)$ & $(s^9_4,0.37)$    \\
$A_3$ & $(s^9_4,-0.35)$ & $(s^9_3,0.25)$  & $(s^9_4,0.45)$  & $(s^9_0,0)$    & $(s^9_4,-0.26)$ \\
\bottomrule  
\end{tabular}%
}
\caption{For each alternative, the LDM4WUE methodology can calculate and report the combined usability scores based on the weighted roles in each test.}
\label{tbl:global}
\end{table}

\subsection{Case \modif{of study -} Phase \modif{5}: exploitation}

TOPSIS was used to rank the alternatives evaluated. The procedure for the two types of ranking achieved in the proposed model is detailed below:

\bigskip
\textbf{Step 12: Generation of rankings by role}. For the following steps, computations are based on the $UCD^l$ matrix for each role $R_l$, as given in Table~\ref{tbl:byRoles}. It should be noted that $UCD^1$ is repeated in Table~\ref{tbl:ucd1}. The positive ideal solution $A^{+}$ and the negative ideal solution $A^{-}$ for $R_l$ are \modif{ determined by Equations \ref{eq:a+l} - \ref{eq:r-l} and represented by $A^{+l}$ and  $A^{-l}$ . For example, we get the following values for $R_1$}:
    $$A^{+1} = \{2.183, 0.439, 1.438, 0.108 \},\\
    A^{-1} = \{1.434, 0.264, 1.067, 0.000 \}$$

The separation measures, $D^{+l}_i$ and $D^{-l}_i$ \modif{for each Role $R_l$}, of each alternative $A_i$ from the positive ideal solutions $A^{+l}$ and the negative ideal solutions $A^{-l}$ are calculated by Equations \ref{eq:d+l}-\ref{eq:d-l} as expressed in Table~\ref{table:resultsG}. Additionally, the relative proximity coefficients $RC^{l}_i (i= 1,2,3)$ are calculated by Equation \ref{eq:rc+l}. These are used to rank alternatives as $Ranking^{l}$ and the results are shown in Table \ref{tbl:byRoles}. 

\begin{table}[h]
\center
\resizebox{\textwidth}{!}{%
\begin{tabular}{llllllcc}
\hline
&
&  \multicolumn{1}{c}{$C_1 \cong SUS$} &
  \multicolumn{1}{c}{$C_2 \cong NPS$} &
  \multicolumn{1}{c}{$C_3 \cong UT$} &
  \multicolumn{1}{c}{$C_4 \cong ACC$} &
  \multirow{2}{*}{\textbf{ucd}$^l_i$} &
  \multicolumn{1}{l}{\multirow{2}{*}{$Ranking^l$}} \\ \cmidrule{3-6}
\multicolumn{1}{c}{} &
  \multicolumn{1}{c}{} &
  \multicolumn{1}{c}{$WC_1 = 0.567$} &
  \multicolumn{1}{c}{$WC_2 = 0.114$} &
  \multicolumn{1}{c}{$WC_3=0.292$} &
  \multicolumn{1}{c}{$WC_4=0.027$} &
   &
  \multicolumn{1}{l}{} \\ \hline
\multirow{3}{*}{$R^1 \cong See$}  & $A_1$  & $(s^9_3,0.45)$  & $(s^9_3,0.31)$  & $(s^9_4,0.34)$  & $(s^9_0,0)$ & $(s^9_4,-0.4)$  & 2 \\
& $A_2$ & $(s^9_4,-0.15)$ & $(s^9_4,-0.15)$ & $(s^9_5,-0.08)$ & $(s^9_4,0)$ & $(s^9_4,0.17)$  & 1 \\
& $A_3$ & $(s^9_3,-0.47)$ & $(s^9_2,0.32)$  & $(s^9_4,-0.34)$ & $(s^9_0,0)$ & $(s^9_3,-0.23)$ & 3 \\ \hline
\multirow{3}{*}{$R^2 \cong Hearing$} & $A_1$ & $(s^9_3,0.22)$  & $(s^9_4,-0.38)$ & $(s^9_4,0.24)$  & $(s^9_0,0)$ & $(s^9_3,0.48)$  & 3 \\
& $A_2$& $(s^9_5,-0.37)$ & $(s^9_5,-0.42)$ & $(s^9_4,-0.34)$ & $(s^9_4,0)$ & $(s^9_4,0.33)$  & 1 \\
& $A_3$& $(s^9_4,-0.12)$ & $(s^9_3,0.13)$  & $(s^9_4,0.39)$  & $(s^9_0,0)$ & $(s^9_4,-0.16)$ & 2 \\ \hline
\multirow{3}{*}{$R^3 \cong Touch$}   & $A_1$ & $(s^9_4,0.02)$  & $(s^9_4,0.18)$  & $(s^9_5,0.07)$  & $(s^9_0,0)$ & $(s^9_4,0.24)$  & 3 \\
& $A_2$ & $(s^9_5,-0.43)$ & $(s^9_5,-0.02)$ & $(s^9_5,0.24)$  & $(s^9_0,0)$ & $(s^9_5,-0.31)$ & 2 \\
& $A_3$ & $(s^9_5,-0.07)$ & $(s^9_5,-0.4)$  & $(s^9_6,-0.43)$ & $(s^9_0,0)$ & $(s^9_5,-0.06)$ & 1 \\ \cline{1-8} 
\end{tabular}%
}
\caption{Ranking of alternatives by roles.}\label{tbl:byRoles}
\end{table}

According to $Ranking^1 = A_2 \succ A_1 \succ A_3$, which focuses on role $R_1=\{\textit{See}\}$, the best usability is shown on website $A_2$ and alternative $A_3$ has the lowest degree of usability for users with this role. In terms of role $R_2=\{\textit{Hearing}\}$, we obtain another perspective as $Ranking^2 = A_2 \succ A_3 \succ A_1$, confirming that $A_2$ has a better usability level over the other two alternatives. Finally, the change in $Ranking^3=A_3 \succ A_2 \succ A_1$ enables us to identify a certain feature in $A_3$ that helped users with $R_3=\{Touch\}$ impairment more than the other two alternatives.

\bigskip
\textbf{Step 13: Global ranking generation}. In order to achieve a global ranking, the $UCD^{l}$ matrices are considered as the basis (see Table \ref{tbl:byRoles}). We then calculate the positive ideal solution $A^{+}$ and the negative ideal solution $A^{-}$ using Equations \ref{eq:a+l}-\ref{eq:r-l} and these are presented below:

    $$A^{+l} = \{2.439, 0.503, 1.354, 0.076\},
    A^{-l} = \{2.015, 0.370, 1.299, 0.000\}$$

Subsequently, the separation measures, $D^{+}_i$ and $D^{-}_i$, of each alternative $A_i$ from the positive ideal solution $A^{+}$ and the negative ideal solution $A^{-}$ are calculated by Equations \ref{eq:d+l}-\ref{eq:d-l}. Similarly, the relative proximity coefficients $RC^{}_i (i= 1,2,3)$ are computed by Equation \ref{eq:rc+l}. All of these results are given in Table \ref{table:resultsG}. 

\begin{table}[h]
\center
\begin{tabular}{rccc|c}
\hline
\multicolumn{1}{c}{\textbf{$A_i$}} & \textbf{$D_i^+$} & \textbf{$D_i^-$} & \textbf{$RC_i$}  & \multicolumn{1}{c}{$Ranking^{global}$} \\ \hline
$A_1$ & 0.440 & 0.053 & 0.108 & 3 \\
$A_2$ & 0.000 & 0.454 & 1.000 & 1 \\
$A_3$ & 0.401 & 0.058 & 0.126 & 2 \\ \hline
\end{tabular}
\caption{Values used to sort the alternatives as in $Ranking^{global}$.} \label{table:resultsG}
\end{table}

According to $Ranking^{global}=A_2 \succ A_1 \succ A_3$, of the three alternative websites, $A_2$ is shown to be the most suitable for real users considering an academic environment.
    
\bigskip
\textbf{Step 14. Retranslation.} We compute the \textit{adjective usability report} ($\textbf{aur}^{l}$) vector for each role $R_l$ and the global \textit{adjective usability report} ($\textbf{aur}^{global}$) vector using Equations \ref{eq:retranslationRoles} and \ref{eq:retranslationGlobal}, respectively. The results are shown in Table \ref{tbl:adjective}. According to the rankings and to the adjective usability reports, $A_2$ is clearly the best choice considering the usability aspects of the interface. Linguistically, the three alternatives have an \textit{OK} score according to adjective SUS labels, but thanks to the proposed \modif{methodology}, we can better understand user experiences at these three sites. 

\begin{table}[!hbt]
\center
\scalebox{0.85}{
\begin{tabular}{@{}ll|cc|ccc@{}}
\toprule
  $A_i$ &
  $R_l$ &
  \textbf{ucd}$^l_{i}$ &
  \textbf{aur}$^l_{i}$ &
  \textbf{ucd}$_{i}$ &
  \textbf{aur}$_{i}$ &
  \textit{Usability}
  \\ 
\midrule
\multirow{3}{*}{$A_1$} & $R^1 \cong See$ &
                  $(s^9_4,-0.4)$ & $(s^{sus}_3,0.20)$ &
  \multirow{3}{*}{$(s^9_4,-0.24)$} &
  \multirow{3}{*}{$(s^{sus}_3,-0.12)$} &
  \multirow{3}{*}{$Ok$} \\
 & $R^2 \cong Hearing$ & $(s^9_3,0.48)$  & $(s^{sus}_2,0.48)$ \\
 & $R^3 \cong Touch$   & $(s^9_4,0.24)$  & $(s^{sus}_3,0.12)$  \\
 \hline
\multirow{3}{*}{$A_2$} &  $R^1 \cong See$ & 
                  $(s^9_4,0.17)$ & $(s^{sus}_3,0.085)$  &
  \multirow{3}{*}{$(s^9_4,0.37)$} &
  \multirow{3}{*}{$(s^{sus}_3,0.185)$} &
  \multirow{3}{*}{$Ok$} \\
 & $R^2 \cong Hearing$ & $(s^9_4,0.33)$  & $(s^{sus}_3,0.165)$  &  \\
 & $R^3 \cong Touch$   & $(s^9_5,-0.31)$ & $(s^{sus}_3,0.345)$ &  \\
\hline
\multirow{3}{*}{$A_3$} & $R^1 \cong See$ &
                  $(s^9_3,-0.23)$ & $(s^{sus}_2,0.23)$ & 
  \multirow{3}{*}{$(s^9_4,-0.26)$} &
  \multirow{3}{*}{$(s^{sus}_3,-0.13)$} &
  \multirow{3}{*}{$Ok$} \\
 & $R^2 \cong Hearing$ & $(s^9_4,-0.16)$ & $(s^{sus}_3,0.42)$ &  \\
 & $R^3 \cong Touch$   & $(s^9_5,-0.06)$ & $(s^{sus}_3,0.47)$ &  \\ 
\bottomrule
\end{tabular}%
}
\caption{Final scores to reporting are given under the $S^{SUS}$ adjective scale.}\label{tbl:adjective}
\end{table}



According to the report provided by the tool, as shown in Figure~\ref{fig:report}, the strength of $A2$ lies in its ability to generate NPS promoters driven by good usability results, in contrast to $A1$ and $A3$, which face accessibility issues and have received the lowest possible ratings. According to the ability to obtain role-based metrics, $A3$ is the highest-rated option when considering users with \emph{touch} impairment role.

\section{Conclusions}\label{sec:conclusiones}

The LDM4WUE methodology integrates linguistic techniques into web usability evaluation and focuses on understanding user roles for improved UX and reduced frustration. We introduce a flexible A/B testing approach that combines user satisfaction assessment with standard tests such as SUS, NPS, UT, and Accessibility reporting. In line with the design thinking principles, we incorporate personas and role-playing to enhance empathy and collaboration in the design process. By considering user knowledge and employing linguistic variables, our proposal aims to balance developer perspectives with user needs, ultimately leading to more effective usability assessments.

In order to encourage its use in real-life design contexts, we have implemented the methodology as an online decision support system (DSS) which streamlines the usability assessment process into five stages: definition of the A/B test; user participation in role-playing; gathering of user information; unification and aggregation of collective data; and the generation of usability feedback reports using the adjective SUS for its closeness to the linguistic decision-making approaches and also for its ease of interpretation. The \modif{USE-AB-DSS} facilitates the creation of new tests and roles, making it user-friendly and accessible for usability engineers. Thanks to the use of a 2-tuple computational linguistic model and various ranking methods, our methodology provides comprehensive insights into areas for improvement, enabling informed decision-making for enhancing IT system usability. For instance, our case of study has evaluated three Moodle platforms under the same conditions in terms of course content and platform settings.

In conclusion, our methodology offers a systematic approach to web usability evaluation and emphasizes the importance of user roles and satisfaction assessment. By incorporating linguistic techniques and design thinking principles, we aim to bridge the gap between developer and user perspectives, leading to more user-centered design solutions. Through practical implementation and application in real-world scenarios, our proposal shows its effectiveness in identifying areas for improvement and enhancing overall user satisfaction with websites. 

Our future work will examine promising avenues for improving A/B testing sensitivity so as to adapt LDM4WUE methodology as suggested in~\cite{Quin24}. We can also analyze the impact that weights have on the criteria to offer an explanation of the ranking~\cite{susmaga24}. This could be included as a new functionality on \modif{USE-AB-DSS} to facilitate new insights into web usability.

\section*{Acknowledgments}
This paper has been supported by the Ministry of Science and Innovation and the Spanish Government under grant number PID2023.150070NB.I00 funded by MICIU/AEI /10.13039/501100011033.

\bibliographystyle{elsarticle-num} 
\bibliography{main}

\end{document}